\documentclass[acmsmall,noacm]{acmart}
\renewcommand\footnotetextcopyrightpermission[1]{} 
\usepackage[T1]{fontenc}%
\usepackage{amsmath}
\usepackage{soul}
\usepackage{supertabular}
\usepackage{multirow}
\usepackage{hhline}
\usepackage{array}
\usepackage{xcolor,colortbl}
\usepackage{ctable}
\usepackage{setspace}
\usepackage{url}
\usepackage{hyperref}
\usepackage{footnotehyper}

\begin{document}

\title{Adversary Models for Mobile Device Authentication}

\author{Ren\'e~Mayrhofer}
\email{rm@ins.jku.at}
\orcid{0000-0003-1566-4646}
\affiliation{%
	\institution{Johannes Kepler University Linz}
	\streetaddress{Altenbergerstr. 69}
	\city{Linz}
	\postcode{4040}
	\country{Austria}
}

\author{Vishwath~Mohan}
\email{vishwath@google.com}
\affiliation{%
	\institution{Google}
	\country{USA}
}

\author{Stephan~Sigg}
\email{stephan.sigg@aalto.fi}
\affiliation{%
	\institution{Aalto University}
	\city{Espoo}
	\country{Finland}
}

\begin{abstract}
Mobile device authentication has been a highly active research topic for over 10 years, with a vast range of methods having been proposed and analyzed. 
In related areas such as secure channel protocols, remote authentication, or desktop user authentication, strong, systematic, and increasingly formal threat models have already been established and are used to qualitatively and quantitatively compare different methods.
Unfortunately, the analysis of mobile device authentication %
is often based on weak adversary models, suggesting overly optimistic results on their respective security.
In this article, we first introduce a new classification of adversaries to better analyze and compare mobile device authentication methods. 
We then apply this classification to a systematic literature survey. %
The survey shows that security is still an afterthought and that most proposed protocols lack a comprehensive security analysis.
Our proposed classification of adversaries provides a strong uniform adversary model that can offer a comparable and transparent classification of security properties in mobile device authentication methods. 
\end{abstract}



\maketitle
\fancyfoot{}
\thispagestyle{empty}

\section{Introduction}
Mobile devices such as smartphones carry an increasing variety of personal data.
For instance, recent proposals to include electronic identities (eID) into smartphones with the aim of replacing classical photo identification documents like driver's licenses or passports~\cite{Hoelzl2017IFIPFull} as well as applications and sensors to more accurately capture the wearer's health status~\cite{hernandez2015biophone}, audible interaction\footnote{https://www.apple.com/ios/siri/}\footnote{https://developer.amazon.com/alexa}\footnote{https://assistant.google.com/} and even emotional state~\cite{likamwa2013moodscope,grunerbl2015smartphone} highlight the breath of sensitive data.

Mobile devices therefore are becoming a critical component in terms of security and privacy not only in the digital domain, but also for interactions in the physical world, with users unlocking their smartphones for short (10-250 seconds) interactions about 50 times per day on average~\cite{hintze2017deviceusage,Falaki_2010}.

In this article, we address \emph{user-to-device} (U2D) authentication, i.e.~users authenticating themselves before being able to use certain functionality of a device~\cite{teh2016survey,patel2016continuous}, as well as two other forms of authentication, \emph{device-to-device} (D2D)~\cite{mayrhofer2013uacap,chong2014survey,fomichev2017securedevicepairing}, and device-to-user (D2U)~\cite{mayrhofer2014architecture,findling2015devicetouser}.

U2D authentication can be performed by one or a combination of four \emph{factors}\footnote{For D2D and D2U authentication, various factors and physical channels have been proposed, but no systematic classification has so far been accepted as common knowledge.}\cite{maciej2019multifactor,ferrag2020authentication}: %
\begin{enumerate}
	\item something a user knows (passwords, PIN codes, graphical patterns, etc.)
	\item something a user possesses (hardware tokens, keys, etc.)
	\item something a user is (static biometric, e.g.\ fingerprint, face, iris, hand geometry, vein patterns)
	\item something a user does (dynamic biometric, e.g.\ gait, handwriting, speech)
\end{enumerate}

Many authentication \emph{methods} have been proposed for mobile devices, but 
not a single established as canonical U2D authentication. 
Instead, approaches have their respective advantages and disadvantages~\cite{hintze2015cormorant,de2015secure}. 
Second factor authentication with something a user possesses often demands D2D authentication through wireless communication. 
Secure D2D authentication is thus a condition to using wireless devices as hardware token for U2D authentication.
In this article, we:
\begin{itemize}
  \item Survey security analysis in the state-of-the-art and derive their assumptions (which are often not explicitly stated) about attackers of such authentication methods. 
  These so-called \emph{adversary models} are a sub-set of threat models commonly applied to cryptographic protocols.
  \item Show that existing security evaluations of these methods often lacks, with many proposals using an insufficient number of subjects or missing independent analysis by others.
  \item Propose a qualitative classification of adversaries to mobile device authentication that enables a more systematic adversary modeling, and use this \emph{scheme} in our review.
\end{itemize}

We design our classification scheme by studying the requirements for useful adversary models at a meta level, with the aim of applying specific instances of these model classes to individual authentication methods. 
Our intention is for this scheme to be used for future research, giving authentication methods a concrete security level to aim for and to test against.
Finally, this article is also a call for action to improve the state of the art in security testing of mobile device authentication.

\section{Authentication on mobile devices}
General threats for user authentication, which include brute-force, password guessing, installing malware, and hardware-level exploits to bypass authentication, typically also apply to mobile devices. %
Mobile device security, however, is inferior to security on desktop computers~\cite{clarke2005authentication,6170530} while mobile devices add additional security threats~\cite{5958024} because of usability issues or limitations due to smaller size, computational and storage capabilities~\cite{botha2009desktop,7905510,NSU2018}. 

\subsection{Adversary models for user authentication}
In a recent meta survey by Ferrag~et\,al.\cite{2018arXiv180310281A} impactful surveys on mobile device authentication are analysed (\cite{6170530,6177188,doi:10.1108/IMCS-03-2013-0019,7000543,6999911,teh2016survey,741bf8a7851144ecb36bf4c5729a5442,patel2016continuous,gandotra2016,8141882,Kunda2018}). 
In total 26 different attacks were described in these publications, 5 threat models and 4 categories of authentication. 
Complementing this work on attacks and countermeasures, we focus on adversary models.

In the literature, formal models are employed to describe the security properties of a particular protocol.
For instance, the most famous adversary model for communication channels is the Dolev-Yao model\footnote{Other types of formal definitions for authentication can be found, for instance, in static analysis or type theory~\cite{bodei2003automatic,abadi1999secrecy}.}, which
assumes that communication partners trust their encrypted messages to the adversary for delivery~\cite{dolev1983security}. %
The adversary can use any information obtained from previous messages to try to decrypt the information.
In particular, she is not constrained by other prior assumptions. %
The Dolev-Yao model is indistinguishable under chosen plaintext attacks, respectively chosen ciphertext attacks (IND-CPA/IND-CCA) for formal cryptographic protocol verification~\cite{bellare1998relations}.

The existence of such accepted standards is crucial for the field since it (1) helps to build trust in security mechanisms, (2) generates a common ground on which approaches are comparable, and (3) creates incentives to build stronger security schemes.
It is also a necessary requirement for (4) the generation of business models grounded on secure technology. 

The literature on mobile device authentication has not yet converged on a common adversary model and while the Dolev-Yao model has been applied for mobile device authentication~\cite{truong2014comparing,shrestha2014drone}, this approach has shortcomings as it does not reflect the specific nature of mobile device authentication.
For instance, mobile devices are typically resource constrained, authentication is conducted in public spaces, potentially under video-surveillance, and the user interface is limited, thus preventing some strong authentication mechanisms.  

The landscape of mobile device authentication methods appears to be fragmented and methods are hardly comparable to each other 
(see~\cite{boyd2013protocols} for a large number of authentication protocols, accompanied with reported attacks).
This creates uncertainty on the security properties and on the appropriateness of any mobile device authentication method. 
Specifically, although recent publications on device authentication bring forward a discussion on potential security threats or attacker studies~\cite{emojiAuth_2017,cha2017boosting,jiang2017smartphone},
a single universally accepted model is lacking. 

Challenges to selecting appropriate security margins and cryptographic parameters are that (1) real-world applications require different security levels~\cite{KSIEZOPOLSKI2007246} and that (2) the resources available on mobile devices place limitations on the authentication method~\cite{burmester2009flyweight}.
Attackers differ in \textit{capability} (ability, training, knowledge), e.g. typical user, developers or manufacturers, and in the \textit{effort} (storage, computational, monetary) they invest, e.g. individuals, organizations or nation states. %
\setlength{\tabcolsep}{2pt}
\ctable[
caption = {Previously proposed adversary models},
label = tab:securityModels,
pos = tbp,
width=1\columnwidth,
doinside=\footnotesize
]{@{}p{3.8cm}p{9.2cm}p{2cm}@{}}{%
}{                          \FL
	\textbf{Paper}            & \textbf{Security model} & \textbf{Year}\ML
Ong et al.~\cite{ong2003quality} & Security levels based on key size, block size, and type of data & 2003\NN
Hager~\cite{hager2004context} & Security levels conditioned on performance, energy and resource consumption & 2004\NN
Ksiezopolski et al.~\cite{KSIEZOPOLSKI2007246} & Different applications require different security levels & 2007 \NN
Gligor~\cite{gligor2007handling} & Adversary model specifically focused on Mobile Ad-hoc networks &2007\NN
Sun et al.~\cite{sun2008quality} & Evaluation model based on quality of protection & 2008\NN
Damg{\aa}rd et al.~\cite{damgaard2008rfid}	& Trade-off between complexity and security in symmetric cryptography & 2008\NN 
Ng et al. and Paise et al.~\cite{ng2008rfid,paise2008mutual} & Security model with adversarial classes based on computational complexity & 2008\NN
Burmester et al.~\cite{burmester2009flyweight} & Mobile device security suffers from limited resources & 2009 \NN
Ahmed et al.~\cite{ahmed2011adaptable} & Security model building on iterative testing of security strength & 2011\NN
Boyd et al.~\cite{boyd2013protocols} & Defines security levels conditioned on key size, block size and type of data & 2013\NN
Do et al.~\cite{do2015forensically} & For forensic investigation, using adversary goals, assumptions and limitations & 2015\NN
Song et al.~\cite{song2015effectiveness} & Metric to measure the strength of pattern lock systems & 2015\NN
Do et al.~\cite{do2015exfiltrating}&Dolev and Yao type of adversary model for Mobile covert data exfiltration&2015\NN
Azfar et al.~\cite{azfar2016android}& Adapt an adversary model for forensic investigations on Mobile phones &2016\NN
Miettinen et al.~\cite{miettinen2018revisiting} & Security levels conditioned on the entropy of the context source & 2018 \NN
Do et al.~\cite{do2019role} & Adversary classified by assumptions, goals and capabilities & 2019\NN
Ferrag et al.~\cite{ferrag2020authentication} &Survey on Threat models for mobile devices& 2020\NN
Hosseinzadeh et al.~\cite{hosseinzadeh2020new}&Adversary model for RFID; grounded on Gong-Needham-Yahalom logic&2020
\LL
}

For this reason, models including attackers of various strength have been proposed~\cite{ng2008rfid,paise2008mutual}. 
For instance,~\cite{ahmed2011adaptable} identify a least-strong attacker by iteratively testing against decreasing security strength. 
Ong~et\,al.\ define security levels based on key size, block size, and type of data~\cite{ong2003quality}.
Instead, Sun~et\,al.\ suggest the quality of protection to measure the level of security~\cite{sun2008quality}, while Hager considers instead performance, energy and resource consumption~\cite{hager2004context}, and Damgard~et\,al.\ analyses the trade-off between complexity and security~\cite{damgaard2008rfid}.
Recently, the analysis of context-based authentication conditioned on the entropy of the context source was proposed~\cite{miettinen2018revisiting}.

Table~\ref{tab:securityModels} summarizes previously proposed adversary models. 
Due to the diversity of mobile devices and applications, a single common adversary model might not be feasible. 
To be useful in general practical application, a meta model, exploiting a \emph{set} of models that account for different usability requirements is needed to qualitatively assess the security level. 
Therefore, in section~\ref{sec:classification-scheme} we introduce a classification scheme for adversary models to support such qualitative comparison.

\subsection{Limitations of traditional authentication schemes}
Electronic devices are traditionally protected via alphanumerical passwords or PIN codes~\cite{furnell2008beyond}.
Due to restrictions in the user interfaces of mobile devices, passwords generate a trade-off between usability and security.
A frequently employed alternative for authentication are graphical patterns.
However, such patterns are vulnerable to shoulder surfing or smudge attacks~\cite{bianchi2011phone,aviv2010smudge}.
In shoulder surfing, the adversary either directly or through video~\cite{yue2014blind} observes the authentication sequence and reproduces it.
In smudge attacks, the adversary visualizes smears on the touch interface left behind as a consequence of user authentication.
The frequent changing of authentication challenges in order to counter these attacks again compromises usability~\cite{dhamija2000deja}.

Alternatively, biometrics, recognising individuals from  behavior and biological characteristics\footnote{International Organization for Standardization. ISO/IEC 2382-37:2012, Information technology -- Vocabulary -- Part 37: Biometrics, 2012}, gained attention for authentication.
This is attributed to biometric sensors included in smartphones, such as fingerprint~\cite{jo2016security} (pore and ridge structure~\cite{stosz1994automated}), voice~\cite{chen2017you} (mel frequency cepstral coding, today deep neural networks~\cite{kaur2020automatic}), gait (heel-strike ratio~\cite{schuermann_2017_bandana} or cycle matching~\cite{muaaz2017imitation}), face (features learned in deep neural networks~\cite{jia2020survey}), keystroke dynamics (key-press latencies~\cite{monaco2015spoofing}), or iris~\cite{wayman2005introduction} (image intensity maps from Hough-transformed Daugman rubber sheet models~\cite{verma2012hough}).

Since biometrics inherit noise, fuzzy pairing is used to account for dissimilarities in the key sequences~\cite{juels1999fuzzy}.
Sequences are mapped onto the key-space of an error correcting code (for instance, BCH or Reed Salomon codes), where $t$ bits can be corrected.
This process also boosts the success probability of an adversary. %
Assuming $|c|$ bit long sequences of which $t$ bits are corrected to result in $|c|-t$ bit long keys, the success probability of a single randomly drawn sequence is then only
\begin{equation}
\sum_{i=0}^t \left(\begin{array}{c}|c|\\i\end{array}\right)/2^{|c|}
=\frac{\sum_{i=0}^t \left(\frac{|c|!}{(|c|-i)! \cdot i!} \right)}{2^{|c|}}.
\end{equation}
\begin{figure}
\centering
\includegraphics[width=.6\columnwidth]{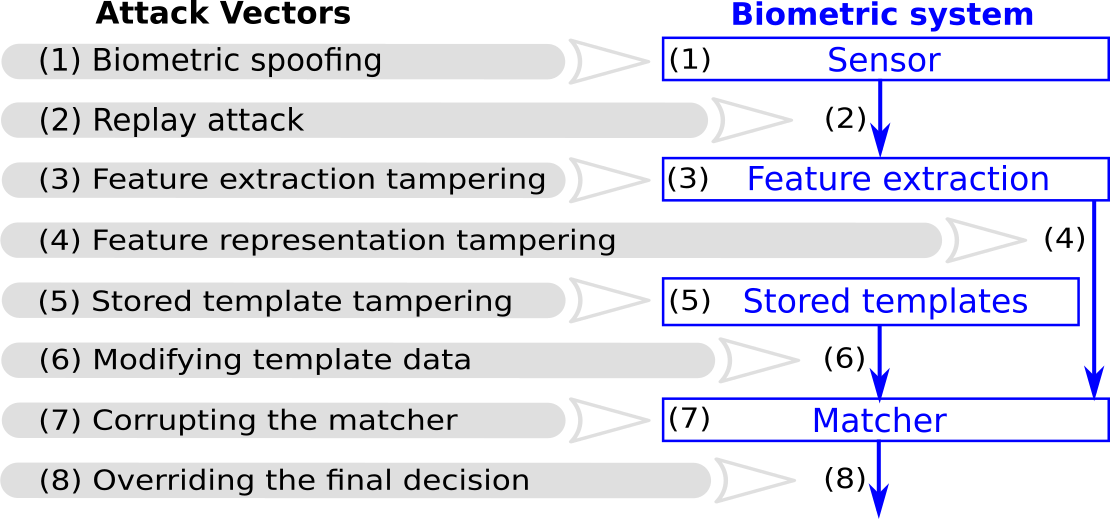}
\caption{Visualisation of generic attack vectors in biometric systems (based on~\cite{ratha2001enhancing})}
\label{figureBiometricAttackVectors}
\end{figure}
However, biometrics can not be kept secret and they can not be revoked~\cite{schneier1999uses,fiebig2014security,schurmann2018moves}.
Consequently, they can not withstand strong attackers under the assumption of targeted spoofing. 

Figure~\ref{figureBiometricAttackVectors} shows attack vectors for biometric systems~\cite{ratha2001enhancing}: 
(1) biometric spoofing~\cite{sharif2016accessorize,Xu_2016_WalkieTalkie,mjaaland2010walk,stang2007gait,kumar_2015_treadmill,muaaz2017imitation}, (2) replay attacks~\cite{gupta2016survey,ruiz2008direct,tey2013can}, tampering with (3) feature extraction~\cite{li2018empirical}, (4) biometric feature representation, (5) stored templates~\cite{stokkenes2016biometric,gafurov2007spoof}, (6) modifying template data~\cite{gupta2016survey,ruiz2008direct}, (7) corrupting the matcher, and (8) overriding the final decision.  
Suggested countermeasures include liveliness detection, supervised enrollment, and securing all stored biometric data~\cite{stokkenes2016biometric}.

\subsection{Adversary models for device-to-device authentication}
Device-to-device authentication is used to pair devices under %
mutual trust.
The information relevant for the pairing can be present at the devices, provided by human ineraction, or acquired from the device's software or hardware sensors.
Also for D2D authentication, capabilities and resources of the adversary are of key relevance for adversary models. 

\begin{figure}
\centering
\includegraphics[width=.6\columnwidth]{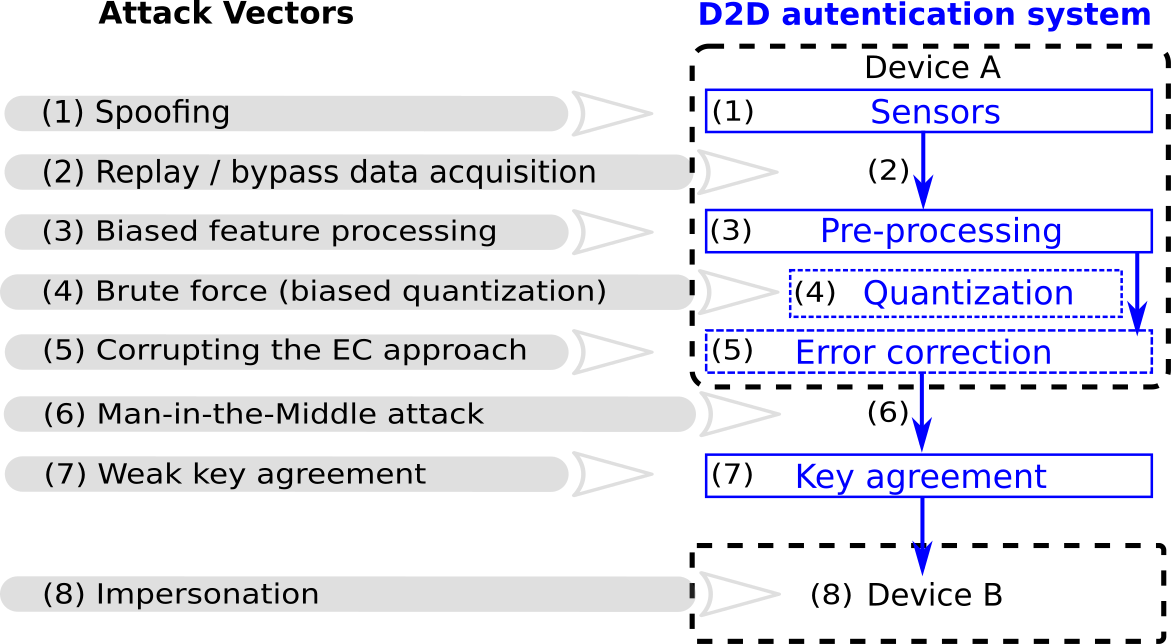}
\caption{Visualisation of generic attack vectors in D2D authentication systems (based on~\cite{bruesch2019secrecy})}
\label{fig:attackVectorsd2d}
\end{figure}

Figure~\ref{fig:attackVectorsd2d} summarizes attack vectors for D2D authentication.
In particular, devices acquire data (stored, human interaction, or sensed), quantize it to bit strings after pre-processing, apply error correction, and agree on a key.
Attack vectors are (1) sensors (forcing the device owner to behave in certain way); (2) bypassing acquisition through replay; (3) biased feature processing; Some protocols employ communication before the actual key agreement~\cite{groza2012saphe,Xu_2016_WalkieTalkie,schuermann_2017_bandana} which might potentially leak information (4);
after error correction, which might be corrupted (5), the key agreement is executed between both devices, thus enabling potential Man-in-the-Middle (MitM) (6), exploitation of weak or false assumption-based key agreement (7) as well as impersonation attacks (8). 

To prevent exhausting the key space, adversaries should be forced into a one-shot model~\cite{vaudenay2005secure}.

For instance, Password Authenticated Key Exchange (PAKE), implemented e.g. by Bluetooth~4.2 Secure Simple Pairing (SSP)\footnote{Because each Bluetooth pairing uses a new ephemeral passkey, SSP does not provide passkey secrecy~\cite{Suomalainen_2007, Sandhya_2012, phan2012analyzing}.}, IPSec, and ZRTP~\cite{Suomalainen_2007,kaufman2014internet,zimmermann2011zrtp}, ensures that the chances of a successful attack depend solely on interactions in the protocol and not on offline computing power~\cite{schmidt2017requirements,mayrhofer2013uacap}.
They thus provide sufficient security margin even with short key sizes K.
Most PAKEs allow for multiple parallel protocol runs~\cite{vaudenay2005secure} and threat models that allow implicit error correction choose a relatively high K = 24 to still have a negligible attacker's success probability even if only 16 out of 24 bits are compared correctly~\cite{farb2013safeslinger}.
Similar margins have been chosen in Bluetooth for PIN comparison with K = 20 and ZRTP for word comparison with K = 20.
Modern PAKEs also provide resilience to dictionary, replay, Unknown Key-Share, and Denning-Sacco attacks~\cite{toorani2014security}, as well as towards mutual authentication, key control, known-key security, and forward secrecy.

Implicit pairing derives secure secrets from similar patterns, e.g. acceleration~\cite{srivastava2015step, heinz2003experimental,findling2014shakeunlock,groza2012saphe}, audio~\cite{schurmann2013secure}, magnetometer~\cite{jin2016magpairing}, or RF features~\cite{mathur2011proximate}, from devices co-present in the same context.

\subsection{Device-to-user authentication (D2U)}
An adversary able to deceive a user into wrongly trusting the identity of a (malicious) device, can harvest user credentials (biometric or knowledge-based) on a subsequent attempt to log in to the device\footnote{This is similar to the well-known credential phishing problem for websites with users mistaking malicious login forms for genuine ones.}.
Device-to-user authentication attempts to address this issue by establishing a means of authenticating the device to the user.
One approach is to visually reveal secret information to establish trust, e.g.\ by displaying variations of secret images to assure authenticity~\cite{roberts2009systems,roberts2010systems}.
However, such systems are prone to shoulder surfing.
Device-to-user authentication is intended to be used frequently and must therefore mitigate usability drawbacks.
Although mutual authentication is well established in D2D authentication (e.g.\ IPsec~\cite{davis2001ipsec}), it is rarely used for authentication involving humans.
One reason for this is that device-to-user authentication is bandwidth limited due to the limited attention span and cognitive resources for the recognition of patterns by the user~\cite{mayrhofer2014architecture}. Initial approaches with vibration patterns have been analyzed~\cite{findling2015devicetouser} but seem impractical so far.

\section{Attacks and modalities}
\label{sec:adversaries}
For all authentication settings (U2D, D2D, D2U) we distinguish various attack types.
An authentication system shall at least prevent \textit{accidental} login from non-authorized users: evaluation against blind guesses (knowledge-based authentication) or samples (biometric)~\cite{govindarajan2013secure,de2012touch,feng2012continuous,frank2013touchalytics}.

Any \emph{targeted} attack will be more powerful~\cite{serwadda2016toward}, for instance, biometric spoofing to exploit weaknesses of specific biometric modalities~\cite{uludag2004attacks,ballard2006biometric}, such as using a picture of a target person in an attempt to spoof face recognition.

An \textit{informed} adversary may also attempt to attack the software implementation~\cite{rahman2013snoop}, or to exploit security breaches in the operating system to leak confidential information about the authentication challenge~\cite{li2013unobservable}, for instance, obtaining 
extraordinary \textit{privileges} to install a keylogger. 

In some cases, historical or other publicly \textit{available data} can be used to elevate chances of a successful attack.
For instance,~\cite{ballard2007forgery} exploits population statistics to launch an attack on a handwriting biometrics system, while~\cite{serwadda2013examining} leverage a typing-database to attack keystroke authentication. %

Adversaries can also \textit{steal} authentication samples to e.g. replay them~\cite{rahman2013snoop}, to train adversaries to forge patterns~\cite{tey2013can,muaaz2017imitation}, or to distort the victim's template and expose it to further attacks~\cite{wang2012transforming}.

The victim sometimes enables attacks through careless actions that lower the effective security (disabling authentication~\cite{frank2013touchalytics}, inadvertently providing access to credentials~\cite{ballard2008towards}).
Mobile systems can counter this by e.g. careful choice of images~\cite{alt2015graphical} or geometric transformations~\cite{schneegass2014smudgesafe}.

Another threat is automated attacks against mobile authentication by \textit{robotic systems}~\cite{serwadda2016toward}.

Finally, \textit{side channels} are a common threat to mobile systems, such as smudge~\cite{aviv2010smudge} and shoulder surfing attacks~\cite{forget2010shoulder,harbach2014sa}. 
Others are the use of on-device accelerometers to recover a PIN~\cite{aviv2012practicality} or infering credentials from channel state information (CSI)\footnote{Changes in electromagnetic signals at a radio receiver caused by movement of a user or object reflecting the signals are visible in the CSI.}~\cite{Li_2016_WiFi}.
Countermeasures include input methods that integrate haptic and audio feedback~\cite{bianchi2011phone}, or applying geometric image transformation~\cite{schneegass2014smudgesafe}.
Another countermeasure is to lower the number of authentication challenges presented by introducing a limited access safe-mode to access non-critical functions without authentication, while falling back to authentication for other functions~\cite{buschek2016snapapp}.

\section{Classifying adversary models}
\label{sec:classification-scheme}
Our classification of adversary models is related to ISO/IEC 62443 security levels in ISA99~\cite{isa99-iec62443-security-levels}\footnote{Only a summary document of this standard is available online at the time of this writing, in the form of public slides by Pierre Cobes; Available online at \url{http://isa99.isa.org/Public/Meetings/Committee/201205-Gaithersburg/ISA-99-Security_Levels_Proposal.pdf}}:

\begin{description}
	\item[SL1] ``Protection against casual or coincidental violation''
	\item[SL2] ``Protection against intentional violation using simple means''
	\item[SL3] ``Protection against intentional violation using sophisticated means''
	\item[SL4] ``Protection against intentional violation using sophisticated means with resources''
\end{description}

In our work, we classify adversaries along two %
dimensions: capabilities and effort (cf. figure~\ref{fig:EffortsResources}).

\emph{Capabilities} in terms of sophistication of specific attacks define an upper bound on the capability of an adversary and which information and secrets they have access to.
This is roughly comparable to the definition of oracles in cryptographic protocol verification.

The \emph{effort} in terms of time, computation, (volatile or non-volatile) memory, and other resources is the upper bound on the amount of energy an adversary is prepared or capable to spend.
This limits the number of trials to attack an authentication method (e.g. the number of guesses in a brute-force attack).

\begin{savenotes}
\begin{figure}
\begin{footnotesize}
 \begin{tabular}{l|m{5.65cm}}
 \multicolumn{ 2}{c}{\textbf{Capabilities}}\\
  \cellcolor{gray!20!white}\textbf{C3} & \cellcolor{gray!20!white} Capabilities of \textbf{manufacturer, owner, operator} (in-depth knowledge; access to cryptographic keys, instructions, or hardware ports (\emph{insider}~\cite{usenix-enigma2019-android-insider-attack-resistance}).\footnote{Within the scope of this article, we do not distinguish between an original manufacturer of a system, the current owner, and a technical operator, but assume the superset of all their capabilities. 
  In terms of cryptographic protocol analysis, this class is most similar to a collusion between all parties besides the actual target system of an attack.}\\
\textbf{C2}   
 & Capabilities of a \textbf{developer} (knowledge about internal structure; no privileged access or possession of cryptographic keys (Kerkhoff's principle)).\footnote{We explicitly do not distinguish between original developers and outsiders, as the internal structure can typically be reversed engineered.}\\
  \cellcolor{gray!20!white}\textbf{C1} & \cellcolor{gray!20!white} 
  Capabilities attributed to an average \textbf{user} (benign user of the system, with no additional knowledge).
 \end{tabular}\hfill
 \begin{tabular}{p{2.5cm}p{2.38cm}p{2.17cm}}
 \multicolumn{3}{c}{\textbf{Effort}}\\
  \cellcolor{gray!20!white}Resources (time, computation, memory, etc.) available to an average \textbf{individual} (dependent on culture, country, time, etc.).
  &
  Resources available to an \textbf{organization} (capture-the-flag teams, organized crime, multinational companies, etc.).%
  & 
  \cellcolor{gray!20!white}Resources available to a \textbf{nation state} (assumed to be able to compel organizations or individuals to assist). %
  \\\hline
  \cellcolor{gray!20!white}\centering\multirow{2}{*}{\textbf{E1}}\\
  &
  \centering\multirow{2}{*}{\textbf{E2}}
  &
  \cellcolor{gray!20!white}\centering\multirow{2}{*}{\textbf{E3}}
 \end{tabular}

\end{footnotesize}
\caption{Adversaries differ with respect to their capabilities and the effort they are prepared to invest}
\label{fig:EffortsResources}
\end{figure}

Both effort and capabilities define qualitative (ordinal) rather than quantitative classes, and the specific ranges will, while distinct, vary with time and context.

Classifying attacks along these dimensions allow systematic comparison of authentication methods with respect to adversary models. 
Practical experience shows an increasing number of attacks with low sophistication (capabilities), but high  computational resources (effort), such as cloud-support or networks of compromised machines)\footnote{In terms of network protocol adversaries, these are often called `script kiddies' or 'bot operators' for DDoS cases.}. 
A preliminary version of our distinction between adversary classes has appeared in~\cite{muaaz2017imitation}.

\end{savenotes}

\begin{figure}
	\centering
	\includegraphics[width=0.4\textwidth]{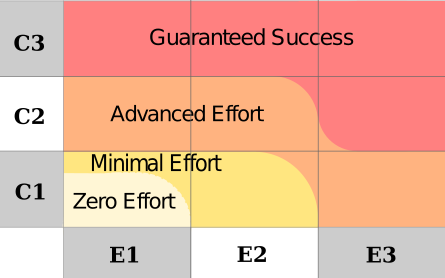}
	\caption{\label{fig:summary-classes}Summary of adversary classes}
\end{figure}

Our scheme assumes that all authentication methods can be broken by an adversary with sufficient capabilities and %
effort. 
However, the required level of capability and effort to do so differs between authentication methods. %
We define in figure~\ref{fig:summary-classes} four classes of adversaries to provide a first, ordinal scale to compare the security efforts different authentication methods have been tested against.

\subsection{Zero effort attacks}
It is common procedure to measure false positive and negative rates for biometric authentication, %
and this is also %
adopted to evaluate authentication schemes such as graphical passwords~\cite{takada2003awase,ritter2013miba,almuairfi2013novel}.
Most quantitative evaluations use $n$ subjects with ground truth and compute the confusion matrix of authentication attempts against stored templates. 
Common measures such as accuracy, precision, recall, true/false positive/negative rates, or F-measure are all based on this same principle%
~\cite{Pervasive_Bishop_2006,bulling2014tutorial}.
We refer to this a zero effort attack because no malicious adversary other than benign subjects exists (adversaries with basic capabilities (\textbf{C1}), and small effort (\textbf{E1})).
Zero effort attacks represent the risk of random success and include na\"ive (non-targeted) brute force attacks.
Examples are honest-but-curious office colleagues, or a stranger who chances upon a misplaced device.

\subsection{Minimal effort attacks}
A minimal effort attack is targeted, e.g.\ mimicing gait, but not with particular sophistication. 
This adversary has no specific system knowledge (\textbf{C1}), but moderate effort (\textbf{E1--E2}). 
Minimal-effort adversaries have the explicit intention of attacking an authentication method and a specific target. 
Many published approaches used minimal effort attacks in their analysis, typically with students or colleagues from the same research group, with %
low effort %
and low to average sophistication.

\subsection{Advanced effort attacks}
Advanced effort attacks show higher sophistication (\textbf{C1-C2}), e.g.\ professional actors trained in imitating body motions, and significant effort (\textbf{E1-E3}). 
We explicitly exclude the combination of (insider) advanced developer-level sophistication with nation-state effort (E3,C2).

\subsection{Guaranteed success attacks}
A guaranteed success attack succeeds in breaking the security of an authentication method.
It allows for any system to describe which capabilities or effort is required for a successful attack.
Authors of device authentication methods are advised to include this adversary class to define the minimum adversary expected to break the system.
An adversary in this class may possess all capability (\textbf{C2-C3}) and effort (\textbf{E1-E3}).
Note that low-effort guaranteed success attacks are possible, for instance through access to cryptographic credentials (E1,C3).

\section{Literature survey: Adversary models for mobile authentication}
We discuss proposals for mobile authentication and adversary models used, and group the literature according to the adversary class utilized to allow a domain-specific discussion of adversary models.
A summary of the publications covered is given in Tables~\ref{tab:literature-survey-zero} to~\ref{tab:literature-survey-guaranteed}. %
We recommend to use the survey as a reference and %
refer the informed reader to the overview in figure~\ref{fig:ToC} to quickly navigate to the section of her interest. 
In addition, attack schemes are collected in tables~\ref{tab:attacks}, \ref{tab:attacks_image}, and \ref{tab:attacks_recall}.
\begin{figure}
\fbox{\begin{minipage}{.97\textwidth}
\begin{scriptsize}
 \begin{tabbing}
\hspace*{0.5cm}\=\hspace*{0.5cm}\=\hspace*{4.6cm}\=\hspace*{1.3cm}\=\hspace*{0.5cm}\=\hspace*{0.5cm}\=\hspace*{0.5cm}\=\hspace*{5cm}\=\kill
\textbf{\ref{sec:biometric} Biometric user-to-device authentication} \> \> \> \pageref{sec:biometric}\> \> \ref{sec:gaze} Gaze-based \> \> \>\pageref{sec:gaze}\\
\> \ref{sec:speech} Speech and audio  \> \>\pageref{sec:speech}\> \> \ref{sec:audio} Audio-based \> \> \> \pageref{sec:audio}\\
\> \ref{sec:keystroke} Keystroke and touch dynamics \> \> \pageref{sec:keystroke}\> \> \ref{sec:acc} Acceleration-based  \> \> \> \pageref{sec:acc}\\
\> \ref{sec:face} Face \> \> \pageref{sec:face}\> \> \>  \textit{Head-movements}  \> \>\pageref{sec:head}\\
\> \ref{sec:iris} Iris \> \> \pageref{sec:iris}\> \> \>  \textit{Acceleration-gestures} \> \>\pageref{sec:gesture}\\
\> \ref{sec:app} Application usage patterns \> \> \pageref{sec:app} \> \textbf{\ref{sec:D2D} Device-to-device authentication} \> \> \> \> \pageref{sec:D2D}\\
\> \ref{sec:gait} Gait \> \> \pageref{sec:gait}\> \> \ref{sec:Dacc} Acceleration-based \> \> \> \pageref{sec:Dacc}\\
\> \ref{sec:finger} Fingerprint \> \> \pageref{sec:finger}\> \> \>  \textit{D2U authentication} \> \> \pageref{sec:D2U}\\
\> \ref{sec:body} Body impedance \> \> \pageref{sec:body}\> \>  \>  \textit{D2D authentication} \> \> \pageref{sec:auth}\\
\textbf{\ref{sec:userToDevice} Usably secure user-to-device authentication}\> \>\>\pageref{sec:userToDevice}\> \> \ref{sec:audio2} Audio \> \>  \> \pageref{sec:audio2}\\
\> \ref{sec:image} Image \> \> \pageref{sec:image}\> \> \ref{sec:token} Token-based \> \> \>\pageref{sec:token}\\
\> \>  \textit{Recognition-based}  \>\pageref{sec:recognition}\> \> \ref{sec:electro} Electromagnetic signals (RF) \> \> \>\pageref{sec:electro}\\
\> \>  \textit{Recall-based} \> \pageref{sec:recall}\> \textbf{\ref{sec:D2U} Device-to-user authentication} \> \> \> \> \pageref{sec:D2U}\\
\> \ref{sec:multi} Multi-touch \> \> \pageref{sec:multi}\> \textbf{\ref{sec:discussion} Discussion on applied adversary classes} \> \> \> \> \pageref{sec:discussion}
 \end{tabbing}
 \end{scriptsize}
 \end{minipage}
 }
\caption{Overview and structure of the literature survey}
\label{fig:ToC}
\end{figure}

\subsection{Biometric user-to-device authentication (bU2D)}\label{sec:biometric}
A large body of work has exploited biometric stimuli for mobile authentication~\cite{meng2015surveying}.
Measures cover e.g.\ spoken audio, keystrokes, face biometrics, gaze, application usage, iris, gait, or fingerprint~\cite{jain2007handbook}.

Most work in this domain show the general feasibility of a working principle (E1,C1; zero effort), by using a small number of subjects distinguished by the modality, but no threat model, attack scenario, or  analysis of password space.
Table~\ref{tab:attacks} summarizes attacks on biometric authentication.

\subsubsection{Speech and audio}\label{sec:speech}
A number of \textit{zero effort} attacks has been considered for biometric systems using speech and audio. 
For instance, in speakersense~\cite{lu2011speakersense}, during a voice phone call a person is identified. 
The system was tested with 17 subjects, achieving 
over 95\% of accuracy. %

However, an adversary actively trying to break the system has not been considered (E1,C1; zero effort). %
For instance, already a \textit{minimal effort} attack using voice impersonation (replay) might trick the system~\cite{chen2017you}. 
In~\cite{chen2017you} a protection against such attack is proposed by exploring the magnetic field emitted from loudspeakers to distinguish between playback and live voice (E1,C2; minimal effort). 
Again, an informed \textit{advanced adversary}, not using magnetics based loudspeakers with access to respecive resources (advanced microphones) could easily circumvent this protection. 

\setlength{\tabcolsep}{2pt}
\ctable[
    caption = {Attacks on mobile biometric authentication systems},
    label = tab:attacks,
    pos = tbp,
    width=1\columnwidth,
    doinside=\footnotesize
]{@{}p{3.1cm}p{1.9cm}p{7.8cm}p{.5cm}@{}}{%
}{                          \FL
\textbf{Paper}            & \textbf{Modality}       & \textbf{Attack scheme} & \textbf{Year}\ML
Gafurov~\cite{gafurov2006robustness} & Gait & Impersonation & 2006\NN
Stang~\cite{stang2007gait}                 & Gait  & Continuous visual feedback impostors & 2007\NN
Gafurov~\cite{gafurov2007spoof} & Gait & Spoof/various attacks & 2007\NN
Ruiz et al.~\cite{ruiz2008direct} & Iris & fake images& 2008\NN
Derawi et al.~\cite{derawi2010unobtrusive}     & Gait       & Active impostor & 2010\NN
Mjaland et al.~\cite{mjaaland2010walk}     & Gait  & Active long-term trained impostors & 2010\NN
Rahman et al.~\cite{rahman2013snoop} & Keystrokes & snoop-forge-replay attack & 2013\NN
Tey et al.~\cite{tey2013can} & Keystrokes & Imitation through Mimesis technique & 2013 \NN
Karapanos et al.~\cite{karapanos2015sound} & Audio & Advanced co-located attackers&2015 \NN
Kumar et al.~\cite{kumar_2015_treadmill}  & Gait  & Treadmill attack        & 2015\NN
Liu et al.~\cite{liu2015snooping} & Keystrokes & Snooping Keystrokes with mm-Audio & 2015\NN
Monaco et al.~\cite{monaco2015spoofing} & Keystrokes & Spoof keypress latencies & 2015\NN
Gupta et al.~\cite{gupta2016survey} & Iris & Attacks:  Masquerade, Replay, Database &2016\NN
Xu et al.~\cite{Xu_2016_WalkieTalkie}     & Gait    & Passive/active impostor (imitation), MitM & 2016\NN
Zhant et al.~\cite{zhang2017hearing} & Speech & Mimicry &2017\NN
Abdelrahman et al.~\cite{abdelrahman2017stay}&Keystrokes&Thermal attacks on mobile user authentication & 2017\NN
Muaaz et al.~\cite{muaaz2017imitation}   & Gait  & Active impostor (imitation) & 2017 %
\NN
Trippel et al.~\cite{kwongyou}            & Gait  & Poisoning acoustic injection attack & 2017\NN
Khan et al.~\cite{khan2018augmented} & Keystroke & Real-time mimicry attack guidance system& 2018\NN
Tolosana et al.~\cite{tolosana2019presentation} & Signature & Analysis of different spoofing (presentation) attacks & 2019\NN
Marcel et al.~\cite{marcel2019handbook} & Biometrics &Handbook of biometric anti-spoofing & 2019\NN
Vyas et al.~\cite{vyas2020preventing} & bio-sensors & Attack types on body area networks using bio sensors& 2020\NN
Tiefenau et al.~\cite{tiefenau2019please} &Biometrics&Attacks bypassing authentication on mobile devices&2020\NN
Neal et al.~\cite{neal2020presentation} &Behaviour&Spoofing (presentation) attacks on various biometrics&2020\NN
Jia et al.~\cite{jia2020face} & Face & Evaluation of 30 face spoofing attacks & 2020\NN
Hagestedt et al.~\cite{hagestedt2020adversarial} & Eye & Attacks on Classifiers for Eye-based User Modelling & 2020
\LL
}
Another example for a system tested only with respect to \textit{minimal effort} attacks is~\cite{zhang2017hearing}. 
The authors utilize the audio system of the phone as a doppler radar to obtain further evidence on speaker identity. 
The authors launched mimicry attacks (adversary with access to video recording and practice), but did not consider advanced (e.g. developer) sophistication.

An example of a comprehensive security discussion in this domain is the usable two-factor authentication based on proximity measured from ambient sound is~\cite{karapanos2015sound}. 
Starting from false acceptance and false rejection rates (\textit{zero effort}), \textit{advanced effort} attacks are considered (similar environment, same media) and the analysis further distinguishes remote from co-located attacks, which then includes \textit{definite success} attacks (E3,C3; guaranteed success). 

\subsubsection{Keystroke and touch dynamics}\label{sec:keystroke}
An overview on the use of keystroke-dynamics for mobile devices is provided in~\cite{teh2016survey}. 
A number of studies consider \textit{zero effort} attacks only, such as~\cite{clarke2007authenticating}, to authenticate phone users via keystroke analysis of their PIN input~\cite{clarke2007authenticating}. 
Authors report equal error rates (E1,C1; zero effort) but ignore active adversaries with access to advanced resources, such as spoofing key-press latencies can be spoofed with a generative keystroke dynamics model~\cite{monaco2015spoofing} via trained replay attacks~\cite{rahman2013snoop} or utilizing audio~\cite{tey2013can} or video~\cite{yue2014blind}.

Examples for studies considering \textit{minimal effort} adversaries targeting specific subjects are~\cite{hwang2009keystroke,bo2013silentsense} or~\cite{de2012touch} to authenticate from dynamics of using pattern-unlock (E1,C1; minimal effort). 
However, all these approaches omit investigation on protocol weaknesses or potential bias in the keystroke dynamics patterns due to statistical distributions over a larger set of users~\cite{serwadda2013examining}.

\subsubsection{Face}\label{sec:face}
Face features for authentication are %
adapted also in commercial hardware\footnote{e.g. Apple FaceID: \url{https://support.apple.com/en-us/HT208108} (An exact adversary model is not publicly documented)}~\cite{samangouei2015attribute,dieter2013towards}. 
An example for a \textit{zero effort} study is~\cite{kim2010person} who test their approach using face, teeth (stereo cameras) and voice on a database with 50 subjects to report EER, FAR and FRR (E1,C1; zero effort), but ignoring targeted attacks using advanced resources such as replay or database attacks. 

Examples for \textit{minimal effort} studies are~\cite{findling2013towards,dieter2013towards} and also~\cite{crouse2015continuous}, who consider to break face-based continuous authentication of 24 subjects by an impostor with no specific system insight (E1,C1; minimal effort). 
These studies were not tested against \textit{advanced attacks}, such as impersonation or dodging via image manipulation~\cite{sharif2016accessorize} or using images from online social networks~\cite{li2018empirical}. 

\subsubsection{Iris}\label{sec:iris}
Iris recognition on mobile phones is constrained by the limited resources of the phone and have been respected in the \textit{zero effort} studdy in~\cite{kim2016empirical}. 
Without an adversary study, only successful instrumentation has been verified (E1,C1; zero effort).  
A number of \textit{advanced effort} attacks against iris verification comprise fake images~\cite{ruiz2008direct}, masquerading (dilation and contact lenses), database and template hacking attacks~\cite{gupta2016survey}.

\subsubsection{Application usage patterns}\label{sec:app}
Application usage patterns constitute another biometric for mobile device authentication, tested, for instance, in a \textit{zero effort} study in~\cite{voris2016you} 
with 50 subjects, but no attack study (E1, C1, zero effort), such as monitoring application usage via other apps on the phone~\cite{sigg2016sovereignty}.  

\subsubsection{Gait}\label{sec:gait}
Despite studies suggesting gait as biometric feature~\cite{rong2007identification,gafurov2007survey,jain2007handbook,derawi2012smartphones}, studies investigating security features and entropy of gait are lacking such as impact of natural gait changes over time by clothing, footwear, walking surface, walking speed and emotion~\cite{nixon1996automatic,boulgouris2005gait,sloman1982gait}.

Early studies on gait-based mobile authentication (shoe-mounted~\cite{huang2007gait,morris2002shoe, bamberg2008gait}; waist-mounted~\cite{rong2007wearable,ailisto2005identifying,casale2012personalization,hoang2013lightweight,hoang2015gait,derawi2010unobtrusive,muaaz2013analysis}; hand, breast pocket, hip pocket~\cite{vildjiounaite2006unobtrusive}) used \textit{zero effort} adversaries, mainly investigating  feasibility (E1,C1; zero effort) and did not consider attacks . 

An example for \textit{minimal effort} studies on gait-authentication are~\cite{gafurov2008performance,gafurov2006robustness}, which consider from pairs of 22 subjects the robustness of gait-authentication against impersonation attacks (E1,C1; minimal effort).   
In an \textit{advanced effort} study in~\cite{muaaz2017imitation}, professional actors were instead employed to mimic the gait of 15 subjects with close physical properties (E2,C2; advanced effort). 
Other \textit{advanced effort} study comprise control of the speed, step-length, thigh lift, hip movement and width of steps~\cite{kumar_2015_treadmill} (E2,C2; advanced effort), intensively training individuals over multiple days~\cite{mjaaland2010walk} (E2,C2; advanced effort) or exploiting a 100+ subject database of gait sequences~\cite{gafurov2008performance,gafurov2007spoof} (E1,C2; advanced effort).
In addition, the high accuracy of video-based gait recognition systems also empowers an adversary to generate a database of gait information on multiple subjects unnoticed~\cite{schurmann2018moves}.  

\subsubsection{Fingerprint}\label{sec:finger}
Biometric authentication using fingerprints is frequently installed in mobile systems~\cite{uludag2004attacks}. 
Typical attack vectors are (1) and (2) in Figure~\ref{fig:summary-classes}, since fingerprint impressions are easily left on surfaces touched~\cite{schneier1999uses}. 
Attacks on Fingerprint-based systems are discussed in~\cite{jo2016security}.
An \textit{advanced effort} study providing countermeasures against such attacks, presents a system combining biometrics, possession, and  continuity features for progressive authentication (switching between different security levels conditioned on the confidence in the authentication)~\cite{riva2012progressive}.
The study comprised 26 attack attempts using 3 attack scenarios in which the attacker had system knowledge and tried to avoid detection via video and audio sensors (E1,C2; advanced effort).

\subsubsection{Body impedance}\label{sec:body}
Rasmussen et al. propose a pulse-based biometric for two-factor or continuous authentication. 
In their approach, a metal keyboard sends small electric current through the user's body of which the frequency response is used for authentication~\cite{martinovic2017authentication}. 
The study investigates usability and disucss the theoretical password space (E1,C1; minimal effort) but lacks a targeted attack study and an investigation on the uniqueness of body impedance in a larger population.

\subsection{Usably secure user-to-device authentication (uU2D)}\label{sec:userToDevice}
Similar to biometrics, some usably secure authentication schemes are conditioned on specific patterns presented for authentication. 
Attacks on these authentication schems, as summarized in table~\ref{tab:attacks_image}, are either related to traditional attacks on authentication systems, or tailored to the respective modality, such as shoulde-surfing, or imitation attacks.
\setlength{\tabcolsep}{2pt}
\ctable[
    caption = {Attacks and security analysis of user-to-device authentication systems},
    label = tab:attacks_image,
    pos = tbp,
    width=1\columnwidth,
    doinside=\footnotesize
]{@{}p{2.3cm}p{2.1cm}p{8.5cm}p{.5cm}@{}}{%
}{                          \FL
\textbf{Paper}            & \textbf{Modality}       & \textbf{Attack scheme} & \textbf{Year}\ML
Dhamila et al.~\cite{dhamija2000deja} & Image & Brute force, observer, intersection attacks & 2000\NN
Thorpe et al.~\cite{thorpe2004towards} & Image & Dictionary attack & 2004\NN
Davis et al.~\cite{davis2004user} & Image & Password distribution & 2004\NN
Ku et al.~\cite{ku2005remote} & Image & Dictionary, Replay, Compromise password file, DoS, predictable $n$, insider & 2005\NN
Dirik et al.~\cite{dirik2007modeling}   & Image  & Dictionary attack & 2007\NN
Hayashi et al.~\cite{hayashi2008use}& Image & Brute force, educated guess, observer, intersection & 2008\NN
Brostoff et al.~\cite{brostoff2010evaluating} & Image & Human bias in password choice & 2010\NN
Sun et al.~\cite{sun2014touchin}   & Multi-touch  & Shoulder-surfing (Video observation and disclosure of exact password) & 2014\NN
Yue et al.~\cite{yue2014blind} & Touch & Technical challenges of blind recognition of touched keys from video & 2014\NN
Huhta et al.~\cite{huhta2015pitfalls} & Acceleration & Attack on the ZEBRA system~\cite{mare2014zebra}, discuss improvements & 2015\NN
Li et al.~\cite{li2016whose} & Acceleration & Imitation of head movement & 2016\NN
Nguyen et al.~\cite{nguyen2016personalized} & Image recall & Shoulder surfing & 2016\NN
Cha et al.~\cite{cha2017boosting} &  Pattern & Optimal conditions for smudge attacks, protection, mitigation strategies & 2017\NN
Zhang et al.~\cite{zhang2017dolphinattack}&Voice& Dolphin attack: Inaudible voice commands & 2017\NN
Kraus et al.~\cite{emojiAuth_2017} & Emoji recall & Shoulder surfing & 2017\NN
Chen et al.~\cite{chen2017you} & Audio/Magnetom. & Machine-based voice impersonation & 2017\NN
Miettinen et al.~\cite{miettinen2018revisiting} & Audio & Impersonation, Man-in-the-Middle & 2018\NN
Prange et al.~\cite{prange2019vision}&various&Threats and design flaws of smart home environments&2019\NN
Prange et al.~\cite{prange2019securing} & various & Model of security incidents with personal items in public; survey and stories & 2019\NN
Shin et al.~\cite{shin2020android} & pattern & Attacks on android pattern lock systems&2020\NN
Alqahtani et al.~\cite{alqahtani2020image} &image &Attacks on machine learning for image-based capcha&2020\NN
Bhana et al.~\cite{bhana2020passphrase} & Various & Usability and security comparison of authentication schemes  & 2020\NN
Vyas et al.~\cite{vyas2020preventing} &Various&Attack prevention schemes for body area networks&2020\LL
}

\subsubsection{Image}\label{sec:image}
Image-based authentication has an advantage~\cite{suo2005graphical,weinshall2004passwords} over password or PIN-based authentication due to improved usability~\cite{wiedenbeck2005passpoints}, and since it is easier to \textit{recognize} or \textit{recall} an image than a text~\cite{dhamija2000deja,de2002vip}.
On the other hand, memorability and security strength of image-based recognition in comparison to PIN and password based solutions when multiple (10-20) of such passwords need to be remembered, has not been considered in the {literature}. %
Davis, et al.~\cite{davis2004user} further found that (1) user password selection is biased by race and gender~\cite{brostoff2010evaluating}, thus lowering password entropy, (2) the need for several rounds to provide a reasonably large password space impairs usability, and (3) recognition-based systems are vulnerable to replay attacks~\cite{tari2006comparison,aviv2010smudge}. 
It is a research challenge to exploit memorability to improve freshness of authentication challenges~\cite{nguyen2016personalized}.

\paragraph{Recognition-based}\label{sec:recognition}
systems condition authenticatoin on the selection of a specific image or groups of images in a partiular order. 
A commercial example is Passfaces\footnote{\url{www.realuser.com}} which uses images from faces for authentication. 
\textit{Zero-effort} studies proposing implementations of this approach with no security investigation, are, for instance~\cite{akula2004image,takada2003awase,de2002vip} (E1,C1; zero effort).

A \textit{minimal effort} study was presented in~\cite{dhamija2000deja}, in which the authors show that failed logins raised to 30-35\% for PIN and password based authentication, while it dropped only to 10\% and 5\% for artwork and photo images. 
Several attacks are discussed (brute-force, observer, intersection), while other attacks, e.g. on the image database or on the system are disregarded (E1,C1; minimal effort). 
Another example for a \textit{minimal effort} study is~\cite{emojiAuth_2017} (replace numerical PIN-pads with emojis), which studied memorability and robustness against shoulder surfing (E1,C1; minimal effort) but did not consider any strong adversaries. 

\paragraph{Recall-based}\label{sec:recall}
graphical password schemes require that a pattern is recalled, e.g. drawing a shape on~\cite{jermyn1999design}.
Since the precision required to establish a sufficiently large password space is high, cued recall schemes provide cues that help to achieve sufficient precision~\cite{blonder1996graphical,birget2006graphical}, such as images to guide the input or distorted and blurred images~\cite{hayashi2008use}.

Example \textit{zero effort} studies are~\cite{de2007passshape} (pure recall) or \cite{pierce2003conceptual,almuairfi2013novel} (cued recall), which focus on usability and the risk of observation attacks (E1,C1; zero effort), but lack an attack study or security analysis. 
 
A \textit{minimal effort} study is~\cite{klein1990foiling} which calculates the size of the password space and remarks that chosen passwords are clustered~\cite{klein1990foiling} (only $10^{-8}$ of the space is used 25\% of the time) (E1,C2; minimal effort).
Other attack vectors, such as shoulder surfing or smudge attacks are not exploited.

In their \textit{adavanced effort} study, Ku et al.~\cite{ku2005remote} study a variation of this schemefor its reparability~\cite{hwang1995reparable}, resistance against dictionary attacks, replay attacks, compromising the password file, denial-of-service, predictable $n$ attack and the insider attack~\cite{bellovin1993augmented,mitchell1996comments,ku2003cryptanalysis,thorpe2004towards} (E2,C2; advanced effort). 
Another example is~\cite{dirik2007modeling}, who analyze the PassPoints scheme (regions in an image constitute an authentication challenge), originally presented in a zero effort study in~\cite{wiedenbeck2005passpoints}.    
The authors present an evaluation approach for graphical password schemes, in which a password consists of a sequence of click points in an image.
For the attack study, the probability of click points was considered as well as attention-related saliency features (luminosity contrast, color contrast, foreground) in a study with more than 100 subjects. 
For the images used, the observed entropy was derived from the observed FoA map (clicking probability to every grid square) (E2,C2; advanced effort). 
Definite success cases and nation state adversaries with strong capabilities are not considered. 

\subsubsection{Multi-touch}\label{sec:multi}
has been proposed to increase the password space, reduce time to input a password and to address security risks through shoulder-surfing and smudge attacks.

Usability issues are in the focus of the discussion while security threats appear as after-thoughts (E1,C1; zero effort)~\cite{ritter2013miba,oakley2012multi}. 
For instance,~\cite{azenkot2012passchords, takada2013extended, takada2014mtapin} propose finger-tapping for multi-touch pin authentication and investigate only usability in their 30-user case study (E1,C1; zero effort). 

In an \textit{advanced effort} study on multi-touch input in~\cite{sun2014touchin}, the authors recruited 30 volunteers to test rotation-invariant multi-touch free-form passwords. 
10 adversaries with access to video recorded password inputs and exact password shapes attacked the system (E2,C1; advanced effort). 

\subsubsection{Gaze-based}\label{sec:gaze}
password entry exploits the movement of the eye for password input~\cite{de2007evaluation}. 

In \textit{zero-effort} usability studies in~\cite{de2007evaluation, weaver2011gaze,kumar2007reducing}, the subjects had to focus on some location on the screen or perform eye gestures (E1,C1; zero effort), whithout any attacker consideration.  

A similar approach was investigated in a \textit{minimal effort} study in~\cite{forget2010shoulder}, where a subject stares for a certain period (the dwell time) at an area on the display to perform an action~\cite{ware1987evaluation}.
The authors evaluated their approach in a user study with 18 subjects and achieved an error rate for the password input of 96\%  (E1,C1; minimal effort), whithout conducting an attack study.

An \textit{advanced effort} study has been prestented in~\cite{bulling2012increasing} and authors investigated the security of gaze-based graphical passwords using saliency masks by theoretical estimation of password space and discussion of threat models (E2,C2; advanced effort).

An example of a medium effort and capability \textit{guaranteed success} study is~\cite{DeLuca:2009:LME:1572532.1572542}. 
Password input by 24 subjects was video-recorded so that attackers could break the system in a single try in 96\% of the cases while the method was robust against simple shoulder surfing (E2,C2;  guaranteed success).

\subsubsection{Audio-based}\label{sec:audio}
A targeted but unsophisticated attack study (\textit{advanced effort}) over audio-based PIN input has been presented in~\cite{bianchi2011phone}. 
Subjects have been instructed and conducted targeted attacks after observing the login process of the target.
However, attackers were artificially limited in their access to the recorded material (e.g. no audio, reduced quality) and time (E2,C2; advanced effort).
In particular, it was derived in~\cite{miettinen2018revisiting} that time is critical in impersonation and Man-in-the-Middle attacks and that otherwise the secure establishing of a shared secret is possible. 

A \textit{guaranteed success} study is presented by~\cite{karapanos2015sound}, who propose to use similarity in ambient audio as a second factor to authentication.
Weak and strong adversaries are considered up to guaranteed success attacks where the adversary is physically located in the same audio context (E3,C3; guaranteed success). 

\subsubsection{Acceleration-based}\label{sec:acc}
\label{sec:gesture}
\label{sec:head}
An example for a \textit{zero effort} attack is~\cite{liu2009uwave}, in which gesture-based authentication from acceleration sequences was investigated for its usability with five subjects (E1,C1; zero effort), without considering an attack study. 

A targeted but unsophisticated attack study (\textit{minimal effort}) is for instance~\cite{li2016whose} (authentication utilising head-movement patterns while listening to an audio pattern).
The authors provided videos of successful authentication attempts to the non-trained amateur attackers to imitate the authentication movements (E2,C1; advanced effort). 

An example of an attack study also covering \textit{guaranteed success} is~\cite{aumi2014airauth}.
An in-air hand gesture authentication system was evaluated through experiments including video-based attacks and allowing to watch the video multiple times, rewind, or to play in slow motion (E2,C2).

\subsection{Device-to-device authentication (D2D)}\label{sec:D2DD2U}\label{sec:D2D}
Device-to-device authentication, or simply pairing typically exploits similarity in context or proximity to achieve seamless authentication~\cite{mayrhofer2007candidate}.
Alternatively, device-to-device authentication can also be realized following the principles of zero-interaction authentication~\cite{corner2002zero}, in which proximity is exploited to verify identity without explicit input, but relying on contextual cues derived from sensor measurements.
In~\cite{truong2014comparing}, security properties of these schemes is evaluated with respect to different sensor modalities and with respect to a Dolev-Yao adversary. 
Table~\ref{tab:attacks_recall} depicts several attacks on device-to-device authentication. %
\setlength{\tabcolsep}{2pt}
{\renewcommand{\arraystretch}{1}
\ctable[
    caption = {Attacks and security analysis of device-to-device authentication systems},
    label = tab:attacks_recall,
    pos = tbp,
    width=1\columnwidth,
    doinside=\footnotesize
]{@{}p{2.5cm}p{1.4cm}p{9cm}p{.5cm}@{}}{%
}{                          \FL
\textbf{Paper}            & \textbf{Modality}       & \textbf{Attack scheme} & \textbf{Year} \ML
Mayrhofer~\cite{mayrhofer2007candidate}   & Arbitrary  & Man-in-the-Middle, DoS & 2007\NN
Schurmann et al.~\cite{schurmann2013secure} 	& Audio & Statistical properties (keys); Brute force, DoS, MitM, Audio amplification attacks & 2013\NN
Truong et al.~\cite{truong2014comparing} & Various & Performance of sensor modalities wrt Dolev-Yao adversary (relay attacks) & 2014\NN
Anand et al.~\cite{anand2016vibreaker} & Audio & Extract vibration sequence from audio noise & 2016\NN
Kwong et al.~\cite{kwongyou} & Acceleration & Active dversary emitting acoustic interference at MEMS resonant frequency. & 2017\NN
Findling et al.~\cite{findling2017shakeunlock} & Shaking & Protocol-specific attacks: observatory, cooperative, handshaking.%
& 2017\NN
Gong et at.~\cite{gong2017piano} & Audio 	& Spoofing, replay and zero effort attacks & 2017\NN
Schurmann et al.~\cite{schurmann2018moves} & Gait & Brute-force, gait mimicry, video, adding a malicious device & 2018\NN
Bruesch et al.~\cite{bruesch2019secrecy} & Gait & Gait-pairing: brute-force, mimicry, video, malicious device, protocol weakness & 2019\NN
Focardi et al.~\cite{focardi2019usable} & QR & Performance, size and security of cryptographic schemes with respect to usability & 2019\NN
Shafi et al.~\cite{shafi2020half} &Spoofing& Attack on the downlink (half-duplex) in cellular communication& 2020
\LL
}
}

\subsubsection{Acceleration-based}\label{sec:Dacc}
\label{sec:auth}

\label{sec:vib}
Examples for \textit{zero effort} studies exploiting vibration are~\cite{lee2018syncvibe,mehrnezhad2014tap,mehrnezhad2015tap}, who exploit shared vibration sequences between physically connected smartphes or physical tapping of devices onto each other. 
The prototypes have been validated for their basic functionality but no attack or user study has been conducted (E1,C1; zero effort).

For this kind of key distribution which utilizes vibration as an out-of-band channel, Anand et al.~\cite{anand2016vibreaker} attack vibration-based pairing schemes by overhearing the audio signature of the vibration pattern (E2,C1; minimal effort).

In an \textit{advanced effort} study,~\cite{van2016accelerometer} authenticate mobile devices towards a remote server, where the challenges are given by the duration of vibration and responses.
A number of security issues is discussed, follwed by a publicly available taxonomy and entropy analysis (E2,C2; advanced effort).

\label{sec:gait2}
Alternatively, gait acceleration has been exploited for authentication between devices which are carried by the same (walking) subject. 
Instantaneous and characteristic variations in the acceleration and gait sequences, that can be extracted at different body positions constitute the features to a pairing key~\cite{lester2004you,cornelius2011recognizing}.  
This problem has been considered in the \textit{zero effort} studies~\cite{muaaz2014orientation,muaaz2015cross,sun2017secure}, which discuss general fasibility, usability such as averse affects of orientation differences as well as cross pocket gait-based authentication (left-to-right) but no adversary study  (E1,C1;  zero effort). 

\textit{Advanced effort} studies on gait-based D2D authentication are, for instance,~\cite{Xu_2016_WalkieTalkie,xu2017gait,revadigar2017accelerometer}, who consider impersonation and man-in-the-middle attacks, passive eavesdropping, impersonation, entropy, randomness, key distribution analysis from a study conducted with 14 subjects analyze the randomness of the resulting key (E1,C2; advanced effort). 
Examples for \textit{guaranteed sucssess} studies on gait-based D2D authentication are~\cite{schurmann2018moves,bruesch2019secrecy}, which concisely compare and evaluate several gait-based D2D authentication protocols, and consider brute-force attacks, gait mimicry, informed attackers that exploit protocol weaknesses, as well as powerful adversaries with access to video or possibility to attach malicious devices unnoticed on the persons body (E3,C2; guaranteed success). 

A further attack on acceleration-based D2D authentication is to actively emit modulated acoustic interference at the resonant frequency of materials in MEMS sensors to control or modify measured acceleration, and thus inject changes to acceleration sequences~\cite{kwongyou}.

\label{sec:shake}
Finally, \textit{minimal effort} studies have been conducted on shaking-based acceleration-pairing~\cite{mayrhofer2007shake,mayrhofer2009shake,liu2014overlapped,groza2012saphe}, where attacker-victim pairs have been built with the purpose of demonstrating the robustness against active attacks (E1,C2; zero to minimal effort).
Attacks on shaking-based pairing protocols are, for instance, investigated in~\cite{findling2017shakeunlock} (observatory, cooperative, handshaking). 

\subsubsection{Audio}\label{sec:audio2}
An \textit{advanced effort} audio-based D2D authentication was proposed in~\cite{gong2017piano} in which devices in proximity (round-trip audio signals) are automatically paired. 
Non-sophisticated replay and spoofing attacks were identified but no attack study conducted (E2,C2; advanced effort).   

A \textit{guaranteed success} study is~\cite{schurmann2013secure}, in which authentication is conditioned on ambient audio.
Statistical properties of the keys are discussed, as well as limitations of the approach and a number of cases in various environments with different noise conditions is considered, also covering definite success attack scenarios where the attacker establishes the same audio context at two distinct places as well as silent cases that would cause the protocol to fail (E1,C2; guaranteed success).
An entropy estimation or adversaries with advanced technical support such as directional antennas have been postponed to later work though.

\subsubsection{Token-based}\label{sec:token}
The authors in~\cite{nguyen2016battery} propose a token-based system to verify user authentication at the time of touch interaction with the capacitive screen of the mobile device.
A \textit{zero effort} usability study is conducted with 12 participants (E1,C1; zero effort).
An example of a \textit{minimal effort} study is~\cite{jin2014magpairing,jin2016magpairing} who exploiting magnetic interaction through a touch screen for token-based implicit two-factor authentication. 
Technical feasibility and theoretical security in comparison to PIN based login are discussed (E1,C2; minimal effort), an advanced and targeted attack study is omitted. 

The \textit{advanced effort} study~\cite{findling2017shakeunlock} proposes token-based mobile-device unlocking over a pre-estaglished secure channel through conjoint shaking. 
Protocol-specific attacks, assuming accelerated knowledge of the adversary were considered (E1,C2; advanced effort).

\subsubsection{Electromagnetic signals (RF)}\label{sec:electro}
Exploiting similarity in physical radio channel characteristics,~\cite{varshavsky2007amigo,mathur2011proximate} consider \textit{advanced effort} attacks using only few subjects.
They consider strong adversaries that might control the radio channel and induce channel fluctuation to bias correlation for devices in proximity (E2,C2; advanced effort).
The adversary has access to historical channel information.
Other advanced attack types, such as beam-tracing simulations are disregarded.

\subsection{Device-to-user authentication}\label{sec:D2U}
As described in~\cite{mayrhofer2014architecture}, an adversary might try to exploit that a device or app is mistaken by the user for another, trusted device or app. 
In this manner, credential information might be derived by the adversary.
This is especially critical when some devices in the usage chain of a mobile service are not physically exposed to the user as, e.g. pointed out for 5G small cell installations in~\cite{vassilakis2016security}.
To protect against such cases, the author of~\cite{roberts2009systems,roberts2010systems} proposes for a user interface (UI) to present a known secret for authentication towards the user. 
This document merely sketches the idea. 
An attack study or even a theoretical analysis of the attack surfaces has not been conducted (E1,C1; zero effort)

A device-to-user authentication approach exploiting vibration patterns has been proposed in~\cite{findling2015devicetouser}.
The authors propose to define specific vibration patterns specifically for a device in order to allow device-to-user authentication. 
The usability has been tested in a study with 12 subjects which targeted on the acceptance of the system.
Patterns have been recognized with 97\% accuracy, however, an attack scenario or adversaries with access to the device or audio in proximity (to potentially reveal the pattern) have not been considered (C1,E1; zero effort).

\setlength{\tabcolsep}{2pt}
\ctable[
    caption = {Selection of publicly available data sets in mobile authentication},
    label = tab:data_sets,
    pos = tbp,
    width=1\columnwidth,
    doinside=\footnotesize
]{@{}p{2.8cm}p{1.4cm}p{8.75cm}p{.5cm}@{}}{%
}{\FL 
\textbf{Paper} & \textbf{Modality} & \textbf{Description} &\textbf{Year}\ML
Gafurov et al.~\cite{gafurov2007spoof} 	& Gait 	& 760 gait sequences from 100 sujects&2007\NN
Liu et al.~\cite{liu2009uwave}   & Acceleration  & $>4000$ gesture samples, $8$~subjects; over multiple weeks; 8~gesture patterns&2009\NN
Fierrez et at.~\cite{fierrez2010biosecurid} & Biometrics &Speech, iris, face, signature, text, fingerprint, hand, keystroke from 400 subjects&2010 \NN
Findling et al.~\cite{dieter2013towards} & Face & 600 high quality, colored 2D stereo vision face images&2013\NN 
Wang et al.~\cite{wang2013retrieval} & Face & Web faces database & 2013 \NN
Galbally et al.\cite{galbally2014probabilistic}&Passwords&KoreLogic dataset of 75,000,000 unique passwords & 2014\NN
Truong et al.~\cite{truong2014comparing} & co-presence & 2303 Samples (co-/non-co-present: 1140/1163); Audio, Bluetooth, GPS, WiFi&2014\NN
Shrestha et al.~\cite{shrestha2014drone} & co-presence & Phone data (temp., gas, humidity, altitude, orient.); 207 samples; 21 locations&2014\NN
Samangouei et al.~\cite{samangouei2015attribute}& Face 	& Database of 152 facial images &2015\NN
Kim et al.~\cite{kim2016empirical} & Iris & 500 iris image sequences from 100 subjects &2016\NN
Costa-Pazo et al.~\cite{Costa-Pazo_BIOSIG2016_2016}& Face & 1190 video sequences of attack attempts to 40 clients & 2016\NN
Patel et al.~\cite{PatelTIFS16} & Face & 9000 (1000 live/8000 spoof) face images & 2016\NN
Ramachandra et al.~\cite{ramachandra2017presentation} & Face & Databases to benchmark presentation attack resilience & 2017\NN
Tolosana et al.~\cite{tolosana2017benchmarking} & Handwriting & e-BioSign signature and handwriting from 65 subjects&2017\NN
Boulkenafet et al.~\cite{OULU_NPU_2017} & Face &4950 real access and presentation attack videos of 55 subjects& 2017\NN
Shrestha et al.~\cite{shrestha2018sensor} & co-presence & 100 audio samples from synchronized audio streams for non-co-present devices &2018\NN
Tolosana et al.~\cite{tolosana2019biotouchpass} & handwriting & e-BioDigit database (on-line handwritten digits) \& benchmark results & 2020
\LL
}

\subsection{Discussion on applied adversary classes}\label{sec:discussion}
The proposed classification of adversary classes has proven useful to distinguish between various approaches in the literature, as summarized in Tables~\ref{tab:literature-survey-zero} to~\ref{tab:literature-survey-guaranteed}.
It is striking that more than half of the literature considered falls into low security classes (\textit{zero effort} or \textit{minimal effort}).
One reason for this is that authors focus on the usability of their approach solely and disregard security.
We suggest that this lax habit need to be broken to develop better authentication approaches.
An insecure authentication approach might be convenient to use, but its usability is low.
Security is also an aspect of usability and must not be taken lightly. 
We should refrain from stressing mostly convenience of usable security approaches.

This picture calls for a need of re-thinking and strengthening attacker models in mobile authentication schemes, and for further research in this direction.
Similarly, benchmark datasets are needed in order to comprehensively compare mobile authentication approaches~\cite{heinz2003experimental}. 
In Table~\ref{tab:data_sets}, we provide a selection of open datasets in mobile authentication. 

\setlength{\tabcolsep}{.1pt}
\ctable[
caption = {Summary Classification of Adversary Models -- zero effort},
label = tab:literature-survey-zero,
pos = tbp,
width=1\columnwidth,
doinside=\scriptsize
]{@{}p{1.1cm}p{1.1cm}p{2.5cm}p{0.35cm}p{0.35cm}p{.9cm}p{0.7cm}p{6.3cm}p{.6cm}@{}}{%
}{                          \FL
	\textbf{Modality}& \textbf{Refer.} & \textbf{ Performance} &  &  & \textbf{\#}& \textbf{Type}& \textbf{Remark} & \textbf{Year} \ML
	Gait & \cite{kale2003gait} & -- & E1 & C1 & 44/25/71& bU2D&Feasibility on 3 gait databases, No attack study& 2003\NN
	Iris		& \cite{kim2016empirical} 	& Detection rate:$99.4\%$ 	& E1 	& C1 & 100 &bU2D& Feasibility and success case. No security analysis & 2016\NN %
	Speech & \cite{revathi2019person} & Rejection rate: $<94.1\%$ & E1 & C1 & 16+44& bU2D&Feasibility study, playback attack resilience, no adversary&2019\NN
	Gait	  & \cite{hoang2013lightweight}& Accuracy: 94.93\% 	& E1 	& C1 &38& bU2D &Android-based gait authentication. No attack study. &2013\NN	
    Gait	  & \cite{hoang2015gait}& FAR: 0\%, FRR: 16.18\% 	& E1 	& C1 &38& bU2D &Naive brute-force success probability; no attack study. &2015\NN
	Gait	  & \cite{ailisto2005identifying}& EER=FAR:6.4\%,FRR:5.4\% 	& E1 	& C1 &36& bU2D &Feasibility of gait for authentication. No attack study. &2005\NN
	Gait & \cite{rong2007identification}& EER: $6.7\%$ & E1 & C1 & 35 & bU2D & Feasibility study, no security discussion & 2007\NN
	Gait	  & \cite{vildjiounaite2006unobtrusive}& EER: 17.2/14.1/14.8\% 	& E1 & C1 &31& bU2D &Gait-authentication from hand/hip-pocket/breast-pocket. &2006\NN
	Gait	  & \cite{rong2007wearable}& EER:5.6\%\&21.1\% 	& E1 & C1 &21& bU2D &Feasibility of gait for authentication.&2007\NN
	Audio		& \cite{lu2011speakersense}	& Accuracy:$>$80\% 	& E1 & C1 & 15+17 & bU2D &Focus on success cases (speaker-distinction) &2011\NN
	Gait	  & \cite{morris2002shoe,bamberg2008gait}& Accuracy:$<97\%$ & E1 	& C1 &15& bU2D &Feasibility of gait (shoe-mounted) for authentication. &2008\NN
	Gait	  & \cite{casale2012personalization}& Accuracy:$<98\%$ 	& E1 	& C1 &10& bU2D &Demonstrate the feasibility of gait for authentication. &2012\NN
	Iris &\cite{mavcek2019mobile}&FAR/FRR:$<6\%/18\%$& E1 & C1 & 10 & uU2D &Feasibility study, general security discussion, no targeted attack & 2019\NN
	Gait	  & \cite{huang2007gait}& Accuracy: 96.133\% 	& E1 & C1 &9 &bU2D &Feasibility of gait (shoe-mounted) for authentication. &2007\NN
	Touch &\cite{dee2019continuous} & accuracy: 100\% & E1 & C1& -- & bU2D &Evaluation details unclear, no adversary study & 2019\NN
	Gait &\cite{gafurov2007survey}& -- & E1 & C1 & -- & bU2D& Discuss security challenges, no attack study & 2007\NN
	Keystroke & \cite{bhana2020passphrase} & -- & E1 & C1 & 112 & uU2D & Entropy and failures for login, no attack study & 2020 \NN
	Pattern		& \cite{de2007passshape} 	& -- 	& E1 	& C1 & 86 & uU2D &shapes of strokes on touch sceen. Questionnaire: Usability& 2007\NN
    Pattern		& \cite{sun2014dissecting} 	& -- 	& E1 	& C1 &81& uU2D &Analytic metric proposed to classify password strength&2014\NN
    Image	  & \cite{de2002vip}& Errors: 2\%--10\% 	& E1 	& C1 &66& uU2D &Errors, usability, password completion time; no attack study. &2002 \NN
    Image	  & \cite{emojiAuth_2017}& Accuracy: 97\% 	& E1 	& C1 &53& uU2D &Human bias in password choice; shoulder surfing robustness. &2017\NN	
    Image	  & \cite{brostoff2010evaluating}& Accuracy: 97\% 	& E1 	& C1 &53& uU2D &Human bias in password choice; no security analysis. &2010\NN
    Gaze		& \cite{forget2010shoulder} & Success rate: 96\% 	& E1 	& C1 & 45 & uU2D &Zero-effort random success study. No security analysis &2010\NN
    Keystroke	& \cite{clarke2007authenticating} &  EER:12.8\%	& E1 	& C1 &32& uU2D  &4-bit and 11-bit pin input; no attack study.&2007\NN
	Touch		& \cite{nguyen2016battery} 	& Accuracy:12\% & E1 	& C1 &12& uU2D &Low energy tokens for interacition with capacitive devices &2016\NN
    Mult.touch	  & \cite{ritter2013miba}& -- 	& E1 	& C1 &30& uU2D &Multi-touch image authentication. No attack study.  &2013\NN
    Image		& \cite{ritter2013miba} 	& --	& E1 	& C1 &30& uU2D  &Focus on usability and password space&2013\NN
	Gaze	  & \cite{de2007evaluation}     & --    & E1 	& C1 &21& uU2D & Usability of 3 eye-gaze methods; General security discussion.&2007\NN
	Image	  & \cite{wiedenbeck2005passpoints}& -- 	& E1 	& C1 &20& uU2D &Improved usability of the PassPoints cued recall scheme.&2005\NN
	QR &\cite{damopoulos2019hands}&Accuracy: $88\%$ & E1 & C1 & 20 & uU2D &Validation and Usability study, no adversaries&2019\NN
	Gaze		& \cite{kumar2007reducing} 	& Error rate: 4\% 	& E1	& C1 & 18 & uU2D &Limited capability threat model (eyes not captured). %
	&2007\NN
	Icons		& \cite{wiedenbeck2006design} 	& Accuracy:$90.35$ 	& E1 	& C1  & 15 & uU2D &Shoulder-surfing robust; focus on usability; &2006\NN
	M.touch		& \cite{azenkot2012passchords} & Entropy: 15.6bits 	& E1 	& C1 & 13 & uU2D &Multi-touch free-form passwords; theoretical password space&2012\NN
	M.touch 	& \cite{oakley2012multi} & -- & E1 & C1 & 10 & uU2D &Success cases and feasibility; no security analysis&2012\NN
	Shaking		& \cite{liu2014overlapped} 	& -- 	& E1 	& C1 & 8 & uU2D &Attack shaking with random acceleration; no video; no entropy&2014\NN
    M.touch		& \cite{takada2013extended,takada2014mtapin} & --  	& E1 	& C1 & 6 & uU2D  &Success cases, usability (memorability \& time); password space&2013\NN
    Radio 		& \cite{conti2017fadewich} 	& TP/FP/FN:$<$95\%/6\%/51	& E1 & C1 & 3 & uU2D &De-authentication method; usability and positive cases&2017\NN
	Image		& \cite{takada2003awase} 	& --   	& E1 	& C1 &--& uU2D &Self-captured images (conceptual study); no security analysis&2003 \NN	
	Image		& \cite{almuairfi2013novel} 	& -- 	& E1 	& C1 &--& uU2D &Implicit authentication by clicking on objects in images.&2013\NN
	Image		& \cite{akula2004image} 	& -- 	& E1 	& C1 &--& uU2D &Images for authentication. Concept; no security analysis&2004\NN
    Image		& \cite{pierce2003conceptual} 	& -- 	& E1 	& C1 &--& uU2D &Image-supported password entry; no security analysis&2003\NN
	Image		& \cite{weinshall2004passwords} & Accuracy:$>$90\% & E1 & C1 &--& uU2D &Acceptance rate up to 5 months after training.&2004\NN
    Image		& \cite{birget2006graphical} & -- & E1 & C1 &--& uU2D &Image-based cued recall; password space and human bias.&2006\NN
    Gesture & \cite{guerar20192gesturepin} & -- & E1 & C1 & -- & uU2D & Secure smartwatch authentication; No security study/analysis &2019\\
    Gaze		& \cite{weaver2011gaze} 	& Success rate: 83\% 	& E1 	& C1 & -- & uU2D &Eye gaze input by clustering gaze points. Only usability&2011 \NN %
	Acceler. &\cite{findling2014shakeunlock} & TPR/TNR:$79\%/86\%$&E1&C1&29&D2D&Non-targeted attacks (random success)&2014\NN
	Gait & \cite{schuermann_2017_bandana} & -- & E1 & C1 & 15 & D2D &Quantization for gait-based pairing. Statistical analysis keys&2017\NN
	Gesture	& \cite{liu2009uwave} 		& Accuracy: 98.6\% 	& E1 	& C1 &5+5& D2D &Authentication from acceleration (DTW matched); usability&2009\NN
	Acceler. &\cite{cornelius2011recognizing}&Accuracy:$85\%$&E1&C1&7&D2D&Feasibility study (correlation); no adversaries&2011\NN
    Gait & \cite{sun2017secure} & Agreement rate:$<89\%$ & E1 & C1 & 5 & D2D &Propose quantization method based on inter-pulse-interval&2017\NN
    Vibration & \cite{lee2018syncvibe} & Success rate: $<$60\% & E1 & C1 & -- & D2D &Common secret via vibration signatures; No security analysis&2018\NN
    Acceler. & \cite{srivastava2015step} & -- & E1 & C1 & -- & D2D & Collocation detection; User study unclear; no adversary & 2015\NN
    Vibration & \cite{findling2015devicetouser} & Success rate:$97.5\%$ & E1 & C1 & 12 & D2U &D2U authentication via vibration patterns; Usability study&2015\NN
    Image/text & \cite{roberts2009systems,roberts2010systems} & -- & E1 & C1 & -- & D2U &D2U authentication via visual cues. &2010
    \LL
    }

\setlength{\tabcolsep}{.1pt}
\ctable[
caption = {Summary Classification of Adversary Models -- minimal effort},
label = tab:literature-survey-minimal,
pos = tbp,
width=1\columnwidth,
doinside=\scriptsize
]{@{}p{1.3cm}p{1cm}p{2.5cm}p{0.4cm}p{0.4cm}p{.75cm}p{0.7cm}p{6.3cm}p{.6cm}@{}}{%
}{                          \FL
	\textbf{Modality}& \textbf{Refer.} & \textbf{ Performance} &  &  & \textbf{\#}& \textbf{Type}& \textbf{Remark} & \textbf{Year}\ML    
	Behavior &\cite{khan2014comparative} & -- & E2 & C1 &40-158& bU2D&Non-sophisticated attacks on multiple biometric systems&2014\NN
	Keystroke & \cite{wang2019improving} & EER: 12\% & E2 & C1 & 104& bU2D & Exploit adversarial noise; Non-targeted attacks & 2019\NN 
    Keystroke   & \cite{alshanketi2019multimodal} & EER: 9.9\% & E2&C1&100&bU2D&Non-targeted comparison of collected keystroke entries & 2019\NN
    Voice		& \cite{zhang2017hearing}& EER:1\%; Accuracy:99\% 	& E2 	& C1 &21& bU2D &Liveness detection system to protect against replay attacks.&2017\NN
    behavior & \cite{mohamed2019challenge} & -- & E2 & C1 & 20 & bU2D & Interactive biometric authentication; non-targeted attack & 2019\NN
    Gait		& \cite{stang2007gait} 	& EER: 26\% 	& E1 	& C2 & 13 & bU2D &Targeted attacks; video-recordings; physical characteristics&2007\NN	
    Face		& \cite{samangouei2015attribute}& Accuracy:$>$0.72 	& E1 	& C1 &152& bU2D &Feasibility; database of 152 images; No security analysis &2015\NN
    Gait		& \cite{derawi2010unobtrusive} & EER: 20.1\% & E1 & C1 & 51 & bU2D &Focus on success cases&2010\NN
	Gait		& \cite{muaaz2013analysis} 	& EER: 22.49\% 	& E1 	& C1 & 51 & bU2D &No attack cases considered &2013\NN
    Face		&  \cite{kim2010person} 	& Error rates:$<$9\% & E1 & C1 &50& bU2D &FAR \& FRR; no dedicated security study &2010\NN
    Handwriting & \cite{tolosana2019biotouchpass}& Mean EER: $14$\% & E1 & C1 & 50 &  bU2D& Non-targeted attack from database of samples & 2020\NN
   	Keystroke	& \cite{de2012touch} 		& Accuracy:$<$57\% 	& E1  	& C1 & 48 & bU2D &Intensive but untargeted (non-sophisticated) attacks &2012\NN 
    Gait		& \cite{hoang2013lightweight} 	& Accuracy:94.93\%	& E1  & C1 & 38 & bU2D &No dedicated attack cases; FAR \& FRR&2013\NN
	Gait		& \cite{muaaz2014orientation} 	& EER:18.965 	& E1 	& C1 & 35 & bU2D &EER for orientation-independent gait authentication.&2014\NN
	Gait/Face  &  \cite{findling2018mobile} & EER: 11.4 \& 5.4 & E2 & C1 & 35 & bU2D & Non-targeted blind matching of patterns between subjects.&2018\NN 
    Gait		& \cite{hoang2015gait} 		& FAR:0; FRR:16.18\%& E1 	& C1 & 34 & bU2D &No dedicated attack cases; FAR \& FRR&2015\NN
    Face		& \cite{dieter2013towards} 	& TP:0.9781; TN:0.9998	& E1 	& C1 &30& bU2D &Focus on positive case &2013\NN
	Keystroke	& \cite{hwang2009keystroke} 	& EER:13\% & E1 	& C1 &25& bU2D &Limited capability attackers: Password provided; pattern not&2009\NN
    Face		& \cite{crouse2015continuous} 	& TP:65\%; FP:35\%& E1 	& C1 &24& bU2D &Victims first interacted with device before handing to impostor&2015 \NN
    Gait		& \cite{gafurov2006robustness} 	& EER: 16\% 	& E1 	& C1 & 22 & bU2D &Active impostor; no matching person height, no actors&2006\NN	
    PPG & \cite{shang2019usable} & acc: 96.31\% & E1  & C1 & 12 & bU2D &Limited capability adversary; brute-force, shoulder surfing & 2019\NN
    Face		& \cite{findling2013towards} 	& TP:93.89; TN:99.95 	& E1 	& C1 &9& bU2D &Positive cases&2013\NN
    Accelerat.	& \cite{mare2014zebra}& Accuracy:85\% 	& E1 	& C2 &20& uU2D &Bracelet: verify typing of legitimate user; weak attacks.&2014\NN		  
	Environm.& \cite{shrestha2018sensor}	& FNR:$<$14.5\% 	& E1 	& C2 & 2 & uU2D &Relay attacks; system knowledge assumed&2018\NN
    Audio		& \cite{chen2017you} 		& FAR/EER/FRR:0/0/<41\% %
    & E1 	& C2 & -- & uU2D &Magnetic field from loudspeakers; No tailored attacks.&2017\NN
    Magnet.		& \cite{jin2014magpairing,jin2016magpairing} & Accuracy: 92\%& E1 & C2 &--& uU2D &Comparison: password space PIN-based login&2016\NN
    Image 	& \cite{jermyn1999design}	& -- 	& E1 	& C2 & -- & uU2D &'Draw a Secret' scheme; theoretical password space; human bias&1999\NN
	Image 	& \cite{thorpe2004towards}	& -- 	& E1 	& C2 & -- & uU2D &Dictionary attacks against graphical password schemes&2004\NN
	Image		& \cite{hayashi2008use} 	& -- & E1 	& C1 &99& uU2D &Usability; Brute force, educated guess, observer, intersection A.&2008\NN
   	App-use		& \cite{voris2016you} 		& -- 	& E1 	& C1 &50& uU2D &Study positive case with professionals &2016\NN
	Headmove	& \cite{li2016whose} 		&  EER$<7\%$,FAR$<5\%$	& E2 & C1 &37& uU2D &Reduced capability video analysis (no audio)&2016\NN
 	Image		& \cite{dhamija2000deja} 	& Success rate:90\%	& E1 	& C1 &20 & uU2D &Discuss possible attacks and countermeasures&2000\NN
 	object &\cite{funk2019lookunlock} & -- & E2 & C1 & 15 & uU2D & HMD auth.; limited capab. attacks, brute force, shoulder surfing & 2019 \NN
    Impedance &\cite{martinovic2017authentication} & Accuracy:$>$87\% & E1 & C1 & 10 & uU2D &User study and theoretical consideration of the password space. &2017\NN
    Drawing		& \cite{sethi2014commitment} 	&  --	& E1	& C1 & 6 & uU2D &Threat model; no attacks; no Entropy; no statistical analysis&2014\NN
    Keystroke	& \cite{bo2013silentsense} 	& Accuracy: 99\%& E1	& C1 &5& uU2D &Random correlation attacks; no sophisticated or active attacker&2013\NN
    Shaking & \cite{groza2012saphe} & Success rate:$<95$\% & E1 & C1 & -- & uU2D &Entropy \& security analysis; no trained, informed adversary&2012\NN
	Shaking		& \cite{bichler2007key} 	& Success rate:80\%	& E1 	& C1 & -- & uU2D &Entropy analysis of the generated keys &2007\NN
	Pattern		& \cite{tari2006comparison} 	& -- 	& E1 	& C1 & 20 & uU2D &Shoulder-surfing robust; low capability, non-trained attack. %
	&2006
	\NN
	Pin		& \cite{roth2004pin} 		& -- 	& E1 	& C1 & 8 & uU2D &Shoulder-surfing robust; complexity analysis; weak attack. %
	&2004\NN
    Audio		& \cite{anand2016vibreaker}	& --& E2 	& C1 &--&D2D&Vibration of devices in contact; extract key from vibration noise.&2016\NN
	Shaking		& \cite{mayrhofer2007shake} 	& FN:10.24\%;FP:0 & E1 	& C2 &8/30/51 & D2D & Competition among limited capability attackers&2007\NN
 	Gait		& \cite{muaaz2015cross} 	& FMR/FNMR:$<$0.09/0.47	& E1 & C1 & 25 & D2D &No attack cases; false non match rate / false match rate (FMR)&2015\NN
    Vibration	& \cite{mehrnezhad2014tap,mehrnezhad2015tap} & FN=FP=EER:9.99\%  & E1 & C1 & 23 & D2D &Synchronized vibration through device-tapping. No attacks.&2014\NN
    Context & \cite{miettinen2018revisiting} & -- & E2 & C1& -- & D2D &Impersonation, MitM, no guaranteed success (same context)& 2018\LL
}

\setlength{\tabcolsep}{.1pt}
\ctable[
caption = {Summary Classification of Adversary Models -- advanced effort},
label = tab:literature-survey-advanced,
pos = tbp,
width=1\columnwidth,
doinside=\scriptsize
]{@{}p{1.3cm}p{1cm}p{2.5cm}p{0.35cm}p{0.35cm}p{.85cm}p{0.7cm}p{6.3cm}p{.6cm}@{}}{%
}{                          \FL
	\textbf{Modality}& \textbf{Refer.} & \textbf{ Performance} &  &  & \textbf{\#}& \textbf{Type}& \textbf{Remark} & \textbf{Year} \ML
    Gait		& \cite{mjaaland2010walk} 	& EER:6.2\%	& E2 	& C2 & 50 & bU2D &Targeted attacks; trained non-professionals; EER: 6.2\%&2010\NN
    Gait		& \cite{kumar_2015_treadmill} 		& FAR:70\% (attack) & E2 	& C2 &18& bU2D &Developer insight (features) \& exploiting treadmill; FAR: 46.66\% &2015\NN
    Behavior & \cite{ku2019draw} &EER/FAR:$<5.9\%/3.3\%$ & E2 & C1 & 30 & bU2D &Public lock pattern, strong attacker with video & 2019\NN
    Face & \cite{jia2020face} & Classification Error rate & E2 & C1& 55/40/40& bU2D&Presentation attacks, targeted from database & 2020\NN
    Gait		& \cite{gafurov2007spoof} 	& EER:13\% 	& E1 	& C2 & 100 & bU2D &Mimicry, non-trained amateurs; non-matching characteristics&2007\NN  
    Behavior & \cite{buriro2019answerauth} & TAR: $99.35\%$ & E2 & C1& 85+6& bU2D &Targeted attack; attacker strength unclear & 2019\NN 
    Fingerprint & \cite{wu2020liveness} & Acc/FAR/FRR:$<99/2/3$ & E2 & C1 & 90 & bU2D& Puppet attack resilience, limited capability (targeted) adversary & 2020 \NN
    Biometric	& \cite{riva2012progressive} 	& Precision= Recall <93\% 	& E1 	& C2 &20& bU2D &System knowledge, audio/video support; 26 attempts; \# unclear&2012\NN
    Image		& \cite{dirik2007modeling} 		& TPR:$>$0.79, TNR:$>$0.68 	& E2 	& C2 &100& uU2D &Automated attack; 70-80\% password points correctly predicted&2007 \NN
    Force & \cite{krombholz2017may} & -- & E2 & C1 & 50+10&uU2D&Targeted attacks, video support&2017\NN
    Pattern & \cite{10.1145/2556288.2557097} & -- & E2 & C2 & 32 & uU2D & Targeted attacks, video support, shoulder surfing& 2014\NN
    Pattern		& \cite{de2013back} 		& Accuracy:44\% 	& E2 	& C2 &24& uU2D & Developer insight \& video analysis (incl. playback)&2013\NN
    Gaze & \cite{fristrom2019free} & TPR/FPR:$81\%/12\%$&E2&C1&29&uU2D&Targeted spoofing attacks on free-form gaze-passwords & 2019\NN
    Image		& \cite{chiasson2007graphical} 	& Success rate:$>$83\%	& E2 	& C2 &24& uU2D & Cued Click Points; shoulder surfing and dictionary attacks&2007\NN
    Gaze/gesture & \cite{abdrabou2019just} & -- & E2 & C2 & 16 & uU2D& Unlimited video access. schoulder surfing resistant & 2019\NN
    Gaze		& \cite{bulling2012increasing} 	& Success. attacks: $<25\%$ & E2 	& C2 & 4+12 & uU2D &Password space; threat models; Attackers with videos&2012\NN
    Pattern		& \cite{cha2017boosting} &  FAR: 74\%	& E2 	& C2 & 12 & uU2D &Smudge attacks: optimal conditions, protection, mitigation&2017\NN
	Audio		& \cite{bianchi2011phone} 	& --  	& E2 	& C2 & 12 & uU2D &Threat: audio-visual recording; Low detail security evaluation  &2011\NN
    Radio 		& \cite{varshavsky2007amigo,mathur2011proximate} & FPR:$<$0.3 & E2 & C2 & -- & uU2D &Access to historical information; MitM; powerful adversary&2011\NN
    Image		& \cite{ku2005remote} 	& -- 	& E2 	& C2 & -- & uU2D &dictionary, replay, pass compromise, DoS, predictable $n$, insider&2005\NN	
	Accel.		& \cite{van2016accelerometer}	& TPR/FPR: 0.7444/0.0978\% 	& E2 	& C2 & -- & uU2D &Security issues; public taxonomy; entropy&2016\NN
	Gaze & \cite{10.1145/3365921.3365922}& & E2&C1&15+25&uU2D&EOG-based, observation-attack resistant, targeted attacks&2019\NN
    Pattern		& \cite{schneegass2014smudgesafe} &  success rate:$>$70\%	& E2 	& C1 & 20 & uU2D &Smudge protection; Limited capability attackers: 3 attempts %
    &2014\NN
    Multi-touch & \cite{sun2014touchin} & TPR:97.5\%, FPR:2.3\% & E2 & C1 & 30& uU2D &Adversary with video \& multi-touch password; FPR: 2.2\%&2014\NN
    Image		& \cite{nguyen2016personalized} 	& -- & E1 	& C2 & 10 & uU2D &Always-fresh auth., Random \& targeted attacks, non-trained,&2016\NN
	Environm.%
	& \cite{shrestha2014drone} 	& FPR:16.25\%, FNR:8.57\% 	& E1 	& C2 &--& uU2D &Adversary with technical understanding of the system&2014   \NN
	Multi-factor & \cite{maciej2019multifactor} & -- & E1 & C2 & -- & uU2D & Replay and MitM; finite automata as adversaries & 2020\NN
    Gait		& \cite{muaaz2017imitation} 		& -- 	& E2 	& C2 &35& D2D & Trained, matched actors, 15 victims; EER: 13\% &2017\NN
	Accelerat. 		& \cite{shrestha2016theft}	& Prec/recall:$>$0.94,0.97	& E2 	& C2 & 20 & D2D &Tap-based pairing via NFC&2016\NN
   Audio 		& \cite{gong2017piano} 		& FRR: 1-12\%; FAR: $<$0.8\% 	& E2 	& C2 & -- & D2D &Co-presence via acoustic signals; spoofing, replay&2017\NN
    Gait & \cite{Xu_2016_WalkieTalkie} & Agreement rate:$<$73\% & E1 & C2 & 20 & D2D &Impersonation, MitM; analyze randomness of keys&2016\NN
	Gait & \cite{xu2017gait} & Agreement rate:$<$73\% & E1 & C2 & 20 & D2D&Impersonation, MitM; analyze randomness of keys&2017\NN
	Gait & \cite{revadigar2017accelerometer} & -- & E1 & C2 & 14 & D2D &Eavesdrop, impersonate, entropy, key randomness/distribution&2017
	\LL
}

\setlength{\tabcolsep}{.1pt}
\ctable[
caption = {Summary Classification of Adversary Models -- guaranteed success},
label = tab:literature-survey-guaranteed,
pos = tbp,
width=1\columnwidth,
doinside=\scriptsize
]{@{}p{1.3cm}p{.9cm}p{2.5cm}p{0.4cm}p{0.4cm}p{.75cm}p{0.7cm}p{6.3cm}p{.6cm}@{}}{%
}{                          \FL
	\textbf{Modality}& \textbf{Refer.} & \textbf{ Performance} &  &  & \textbf{\#}& \textbf{Type}& \textbf{Remark} & \textbf{Year} \ML
	Audio 		& \cite{karapanos2015sound}	& FAR=FRR=$<0.01$ 	& E3 	& C3 & 32 & uU2D &Incl. attakcer in same context (guaranteed success).&2015\NN
   	Generic		& \cite{wu1998secure} 		& -- 	& E3 	& C3 & -- & uU2D &Theoretical study: protocol security, various attacks&1998\NN	
	Gaze        & \cite{DeLuca:2009:LME:1572532.1572542} & FAR:42\% (attack) & E2 & C2 & 24& uU2D &Shapes with gaze. video breaks system, shoulder surfing not&2009\NN
    Pattern		& \cite{kwon2014tinylock} 	& -- 	& E2 	& C1 & 24 & uU2D  &Smudge protection; image analysis to detect smudge patterns&2014\NN
	Magnet.		& \cite{aumi2014airauth} 	& EER:96.6\% 	& E2 	& C2 & 10-15 & uU2D &sophisticated attackers, video analysis, (slow motion, rewind)&2014\NN
	Pattern		& \cite{aviv2010smudge} 	& -- & E2 	& C1 & -- & uU2D &Smudge protection; Case study needs further detail&2010\NN
	Pattern		& \cite{von2013making} 		& Error rate:9.5\%	& E1 	& C2 & 24 & uU2D &Smudge protection; Single security expert attacker&2013\NN
	Gait &\cite{bruesch2019secrecy} & -- & E3 & C3 & 15+482 &D2D &Brute-force, mimicry, video, malicious device, protocol flaws&2019\NN
	Vibration 	& \cite{liu2017vibwrite}	& Accu/FPR:$>$95\%/$<$3\%	& E3 	& C3 & 15 & D2D &Implicit vibration an surface. Different attacker classes&2017\NN
	Audio, light 	& \cite{shrestha2018sensor}	& -- 	& E3 	& C3 & -- & D2D &Proximity detection; active adversaries; 538 audio samples&2018\NN
	Gait &\cite{schurmann2018moves} & -- & E3 & C2 & 15+482 & D2D &Brute-force, mimicry, video, malicious device&2018\NN	
	EMG 		& \cite{yang2016secret}		& Bit mismatch rate:$<$0.4 %
    & E3 	& C2 & 10 & D2D &EMG signals for device pairing &2016\NN
    Various 	& \cite{truong2014comparing}	& FPR:$<27\%$	& E3 	& C2 & -- & D2D &Co-presence: WiFi, GPS, Bluetooth, audio; strong adversary &2014\NN
    Audio, light		& \cite{miettinen2014context} 	& -- 	& E2 	& C2 &--& D2D &Context-based pairing; replay, same-context definite success %
    &2014\NN
	Shaking		& \cite{mayrhofer2007candidate} & -- & E2 & C2 & -- & D2D &MitM, DoS, incl. definite success (low noise channel)&2007\NN	
    Shaking		& \cite{findling2017shakeunlock} & EER:0.1293 	& E1 	& C2 & 29 & D2D &Observatory, cooperative, handshaking. No video or entropy %
	&2017\NN    
	Audio		& \cite{schurmann2013secure} 	& -- 	& E1 & C2 & 2 & D2D &Statistical key properties; attacks incl. guaranteed success&2013\LL
}

\section{Conclusions}
Too many publications use weak adversary models, which is comparable to the early work on cryptography, notably symmetric ciphers.
In cryptography, nowadays, %
new cipher proposals are only considered secure candidates\footnote{They will never be more than candidates. After all, there is typically no formal proof of security, only the absence of specific attacks that mark a `secure' cipher even to this day.} after many other, typically more capable, cryptographers have tried to break it.

We recommend to adopt this attitude for research on authentication methods and, in particular, in the domain of mobile/ubiquitous/wearable/embedded devices.
Studies often mix usability and security concerns, which is commendable, because security is an important aspect of usability. 
However, security is often considered as an after-thought and employing non-security experts as participants can only provide an estimate for false negatives, but have little validity for the false positives of an authentication system. 
To estimate these false positives, a class of significantly stronger attackers is required.

Authors should use realistic attacker models of adversaries who have a real motivation in breaking the system and who are potentially either more skilled than the average user of the system and/or willing to spend significantly more effort than a legitimate user (i.e.\ false matches are allowed much more effort than true matches).
For a comprehensive discussion of a new model -- and new authentication papers should go this far in their own evaluation -- authors should use additional attacker models that will indeed break that authentication method.

The boundary of what security a system can achieve lies between \textit{advanced effort} and \textit{guaranteed success} categories, i.e.\ how far it is capable of providing protection against targeted attacks.
Good authentication methods should define this security level as precisely as possible.

\section*{Acknowledgments}
This work has partially been supported by the LIT Secure and Correct Systems Lab funded by the State of Upper Austria.
All authors had equal contribution and are listed in alphabetic order.

\bibliographystyle{ACM-Reference-Format}
\bibliography{literature}

\vfill

\end{document}